\documentclass[preprint,amsmath,amssymb,aps]{revtex4-1}

\usepackage{graphicx} 
\usepackage{dcolumn} 
\usepackage{bm} 
\usepackage{here}
\usepackage{color}
\usepackage{amsmath,amsthm,amssymb,cases}

\begin{document}

\title{Inelastic neutron scattering of hydrogen in palladium studied by semiclassical dynamics}

\author{Motoyuki Shiga}
\email{shiga.motoyuki@jaea.go.jp}
\affiliation{Center for Computational Science and e-Systems, Japan Atomic Energy Agency, Chiba 277-0871, Japan}

\author{Bo Thomsen}
\affiliation{Center for Computational Science and e-Systems, Japan Atomic Energy Agency, Chiba 277-0871, Japan}

\author{Hajime Kimizuka}
\affiliation{Department of Materials Design Innovation Engineering, Nagoya University, Aichi 464-8603, Japan}

\date{\today}

\begin{abstract}

Inelastic neutron scattering (INS) spectra of hydrogen in face-centered cubic palladium have been calculated considering nuclear quantum effects (NQEs) at finite temperatures.
The calculations were performed using semiclassical Brownian chain molecular dynamics (MD) [Shiga, J. Comput. Chem. {\bf 43}, 1864 (2022)] and artificial neural network potentials with an accuracy of generalized gradient approximation of density functional theory.
The calculated spectra are in good agreement with experimental spectra with respect to the peak positions and intensities corresponding to the fundamental tone and the first overtone of the vibrational excitation of hydrogen atoms.
These results differ significantly from those of classical MD, indicating that NQE plays an essential role in the correct estimation of the INS spectrum.
Importantly, the NQE acts as a blue-shift of the INS spectrum for hydrogen in the octahedral site,
 due to strong anharmonic vibrations of hydrogen on the potential surface with even symmetry.
The calculated peak shifts associated with Pd lattice distortion were also in agreement with experimental results.

\end{abstract}

\maketitle

\section{Introduction}

The physical behavior of hydrogen in metals has long been a topic of basic and applied science, providing a wealth of fundamental knowledge about hydrogen storage and hydrogenation catalyst materials for clean energy technologies \cite{alefeld1978hydrogen,fukai2006metal}.
Metal palladium (Pd) is considered a unique material with a strong affinity to hydrogen because of both its catalytic and hydrogen absorbing properties \cite{adams2011role}.
Inelastic neutron scattering (INS) is a powerful experimental approach that can detect hydrogen in condensed phases, providing valuable information about the stable sites and vibrational motion of hydrogen atoms in Pd crystals and nanoparticles
\cite{bergsma1960thermal,chowdhury1973neutron,drexel1976motions,howard1978vibrational,rush1984direct,nicol1987neutron,nicol1988inelastic,kolesnikov1991neutron,nakai1992neutron,stuhr1995vibrational,ross1998strong,kemali2000inelastic,heuser2008vibrational,heuser2011small,ju2011comparison,heuser2014direct,kofu2016hydrogen,kofu2017vibrational,kofu2020dynamics,otomo2020structural,antonov2022lattice}.


Much effort has been devoted to the theoretical interpretation of the experimental INS spectra of hydrogen in Pd
\cite{rahman1976phonon,gillan1986simulation,salomons1990lattice,elsasser1991vibrational,li1992molecular,trinkle2011nanoscale,errea2013first,paulatto2015first}.
Anharmonic effects of hydrogen vibration are believed to play an important role \cite{elsasser1991vibrational,errea2013first,paulatto2015first}.
Thus, molecular dynamics (MD) simulation is a promising approach fully considering the anharmonicity of the potential energy surface (PES)
\cite{rahman1976phonon,salomons1990lattice,li1992molecular}.
On the other hand, the accuracy of the PES of hydrogen in Pd from {\it ab initio} density functional theory (DFT) has been improved over the years \cite{caputo2003h,ishimoto2018theoretical}.
Conventional MD methods should be reliable for high temperatures if they are performed with an accurate potential model that reproduces the {\it ab initio} calculations.
For low temperatures, however, advanced MD techniques are required to reflect the quantum behavior of lightweight hydrogen atoms, such as zero-point vibrations and tunneling 
\cite{gillan1986simulation}.
In fact, in this paper, we show that the spectrum from conventional MD simulation differ significantly from the experimental spectrum at room temperature.
This is because the anharmonic nature of hydrogen vibrations is underestimated in the absence of nuclear quantum effects (NQEs).
Alternatively, the quantum states of anharmonic hydrogen vibrations can be solved by the Schr\"odinger equation under the approximation of coupling to Pd phonons
\cite{elsasser1992first,kemali2000inelastic,errea2013first,paulatto2015first,ozawa2023observation}.
While this approach is useful, the change in INS spectral shape and its temperature dependence observed in the experiment cannot be reproduced.


Path integral MD (PIMD) \cite{parrinello1984study} and path integral hybrid Monte Carlo (PIHMC) \cite{tuckerman1993efficient} methods are useful for computing quantum statistical ensembles of complex many-body systems at finite temperatures.
Based on the imaginary time path integral formulation of quantum statistical mechanics 
 \cite{feynman1972statistical,feynman2010quantum,schulman2012techniques},
the fact that the quantum fluctuations of a given particle are equivalent to the fluctuations of an interconnected classical replica via harmonic chain allows rigorous computation of time-independent quantum statistics \cite{chandler1981exploiting}.
PIMD simulations are an established technique for exploring NQEs of a wide range of hydrogen-based materials
\cite{marx2009ab,tuckerman2010statistical,shiga2018path,markland2018nuclear,thomsen2022structures}.
On the other hand, rigorous computation of time-dependent dynamical quantum properties is difficult except for simple or few-body systems.
This is because the inclusion of quantum phase according to real-time path integral theory makes numerical calculations unstable due to the oscillatory behavior of complex functions.
To ensure numerical stability while maintaining a balance between computational accuracy and efficiency, a semiclassical approximation must be introduced.


From various semiclassical approximations \cite{cao1994formulation-2,craig2004quantum,krajewski2004quantum,rossi2014remove,liu2014path,hele2015boltzmann,cendagorta2018open, trenins2019path,kapil2020inexpensive, hasegawa2023nuclear,shiga2022path}, we choose to employ the Brownian chain MD (BCMD) method that has  recently been proposed \cite{shiga2022path}.
The BCMD method is akin to centroid MD (CMD) \cite{cao1994formulation-2} and ring polymer MD (RPMD) \cite{craig2004quantum} methods, which proved their success in describing NQEs on hydrogen diffusion in metals
\cite{kimizuka2018mechanism,kimizuka2019unraveling,kimizuka2021two,kwon2023accurate}.
The BCMD method is an extension of the PIMD and PIHMC methods, and has in common with CMD and RPMD its rigorous treatment of quantum statistics at thermal equilibrium.
The BCMD method has been designed to improve the accuracy of the calculations of vibrational spectra, which is known as a crucial problem for the CMD and RPMD methods \cite{witt2009applicability,ivanov2010artificial}.
The quantum canonical correlation function (Kubo-transformed correlation function) of the BCMD method meets the basic physical requirements, such as being accurate in short time propagation, preserving time symmetry, satisfying fundamental conservation laws, having the correct high-temperature/classical limit, and giving accurate results for the position and velocity autocorrelations of harmonic oscillators.
Furthermore, the time evolution of the BCMD is guaranteed to maintain thermal equilibrium and not suffer from zero-point energy leakage problems.
So far, however, the BCMD method has proved successful only in the calculation of infrared vibrational spectra of molecular systems such as liquid water.


In this paper, the BCMD method was used to calculate the INS spectra of hydrogen in face-centered cubic (fcc) Pd.
The NQE of the INS spectra was considered by a semiclassical approximation of the quantum canonical correlation function at finite temperature.
This is the first application of the BCMD method to the vibrational properties of solid systems.
An artificial neural network (ANN) potential was used to ensure the accuracy of the Born-Oppenheimer PES of DFT\@.
The semiclassical BCMD results were shown to be a significant improvement over the classical MD results when comparing the spectral shapes from calculation and the INS experiments, for both the fundamental tone and overtone signals.
Interestingly, the NQE of hydrogen vibrations at the most stable octahedral (O) site of Pd acts as a spectral blue-shift, unlike the red-shifts that are observed in molecular systems.

\section{Theory}

\subsection{BCMD}

We consider a system of $N$ distinguishable atoms whose motion is described
by the Hamiltonian,
\begin{eqnarray}
 {\hat H} = \sum_{I=1}^N \frac{\hat{\bf P}_I^{2}}
 {2M_I^{}} + V(\hat{\bf R}_1^{},\cdots,\hat{\bf R}_N^{}),
\end{eqnarray}
where $\hat{\bf R}_I^{}$, $\hat{\bf P}_I^{}$, and $M_I^{}$ are the position operator,
 the momentum operator, and the mass of atom $I$, respectively.
The path integral expression of partition function of the quantum canonical ensemble of this system is given by
\begin{eqnarray}
 Z &=& {\rm Tr} \left( e^{-\beta {\hat H}} \right)
 \nonumber \\ &=&
 \lim_{P\rightarrow\infty} \prod_{I=1}^N \left[ 
 \left( \frac{M_I^{}P}{2\pi\beta\hbar^2} \right)^{\frac{3P}{2}}
 \prod_{s=1}^P \int d{\bf R}_I^{(s)} \right]
 {\rm e}^{- \beta V_{\rm eff}[{\bf R}] },
 \nonumber \\
 \label{eq2}
\end{eqnarray}
where $\beta = \frac{1}{k_{\rm B}^{}T}$ with the Boltzmann constant $k_{\rm B}^{}$ and the temperature $T$,
$\left\{{\bf R}_I^{(1)}, \cdots, {\bf R}_I^{(P)}\right\}$ is the position of atom $I$
along the imaginary time from 0 to $\beta\hbar$, and $P$ is the number of imaginary time slices or ``{\it beads}''.
Equation (\ref{eq2}) is proportional to the
 classical partition function of a $P$ replicated system, in which the effective potential is of the form
\begin{eqnarray}
 V_{\rm eff}^{}[{\bf R}] &=&
 \sum_{I=1}^N \sum_{s=1}^P \frac{M_I^{} P}{2\beta^2\hbar^2}
 \left({\bf R}_I^{(s+1)}-{\bf R}_I^{(s)}\right)^2
 \nonumber \\
 &+& \frac{1}{P} \sum_{s=1}^P
 V\left({\bf R}_1^{(s)},\cdots,{\bf R}_N^{(s)}\right).
 \label{eq3}
\end{eqnarray}
Now we introduce a linear transformation {(so called the ``{\it normal mode coordinates}'' of beads)} of the atomic coordinate ${\bf R}_I^{(s)}$ as
\begin{equation}
 {\bf Q}_I^{(\alpha)} =
 \frac{1}{\sqrt{P}} \sum_{s=1}^P U_{s \alpha}^{} {\bf R}_I^{(s)},
 \label{eq4}
\end{equation}
such that the first term of on the right-hand side of Eq. (\ref{eq3}) is diagonalized,
where $(U_{1\alpha}^{},\cdots,U_{P\alpha}^{})$ is the corresponding eigenvector of the $\alpha$-th mode.
Then Eq. (\ref{eq3}) can be rewritten as
\begin{equation}
 V_{\rm eff}^{} = \sum_{I=1}^N \sum_{\alpha=2}^P
 \frac{M_I^{}P\lambda^{(\alpha)}}{2\beta^2\hbar^2}
 {{\bf Q}_I^{(\alpha)}}^{2} + \frac{1}{P}
 \sum_{s=1}^P V\left({\bf R}^{(s)}(\{{\bf Q}\})\right),
 \label{eq5}
\end{equation}
where the eigenvalues associated with the eigenvector $U_{s \alpha}^{}$ for $1\le s\le P$ are
\begin{equation}
 \lambda^{(2k)} = \lambda^{(2k+1)} = 4P \sin^2\left(\frac{\pi k}{P}\right)
 \ \ \ \left(1 \le k \le \frac{P}{2}\right).
 \label{eq6}
\end{equation}


Now, BCMD is the combination between a Newton-like equation for the centroid coordinates [${\bf Q}_I^{(1)} = \frac{1}{P} \sum_{s=1}^P {\bf R}_I^{(s)}$],
\begin{equation}
 M_I^{} \ddot{\bf Q}_I^{(1)} =
 - \frac{\partial V_{\rm eff}^{}}{\partial {\bf Q}^{(1)}},
 \label{eq7}
\end{equation}
and an overdamped Langevin equation for the non-centroid coordinates [${\bf Q}_I^{(\alpha)}$ for $\alpha\ne 1$],
\begin{equation}
 M_I^{} \gamma^{(\alpha)} \dot{\bf Q}_I^{(\alpha)} =
 - \frac{\partial {V}_{\rm eff}^{}}{\partial {\bf Q}_I^{(\alpha)}}
 + \sqrt{\frac{2 M_I^{}\gamma^{(\alpha)}}{\beta}} \boldsymbol{\zeta}(t),
 \label{eq8}
\end{equation}
where $\boldsymbol{\zeta}(t)$ is the white noise.
The friction parameter is set as
\begin{equation}
 \gamma^{(\alpha)} = \frac{P \lambda^{(\alpha)} }{\tau^{(\alpha)}}
 = \frac{P \lambda^{(\alpha)} }{\beta\hbar},
 \label{eq9}
\end{equation}
which looses memory in the decoherence time scale of a quantum free particle.
Unlike in CMD and RPMD, the non-centroid motion in BCMD follows a first-order stochastic differential equation, i.e., Eq. (\ref{eq8}).
This corrects the unphysical shifts and resonances in the vibration spectra that are problematic in CMD and RPMD \cite{shiga2022path}.


Finally, time correlation function,
\begin{equation}
 \tilde{C}_{AB}^{\rm bcmd}(t) =
 \langle \overline{A}(0) \overline{B}(t) \rangle_{\rm bcmd}^{},
 \label{eq10}
\end{equation}
 is computed with respect to the bead averages
 of the BCMD trajectory as
\begin{equation}
 \overline{X}(t) = \frac{1}{P}
 \sum_{s=1}^P X\left({\bf R}^{(s)}(t),{\bf P}^{(s)}(t)\right).
 \label{eq11}
\end{equation}
Equation (\ref{eq10}) regarded as an approximation of quantum canonical correlation function,
\begin{equation}
 \tilde{C}_{AB}^{}(t) = \frac{
 \frac{1}{\beta\hbar} \int_0^{\beta\hbar} {\rm d}\tau {\rm Tr} \left( {\rm e}^{-\beta{\hat H}} {\hat A} {\rm e}^{{\rm i}{\hat H}(t+{\rm i}\tau)} {\hat B} {\rm e}^{-{\rm i}{\hat H}(t+{\rm i}\tau)} \right) }
 { {\rm Tr} \left({\rm e}^{-\beta{\hat H}}\right) }.
 \label{eq12}
\end{equation}

\subsection{Dynamic structure factor}

We consider an atomic scattering process with momentum transfer
 $\hbar{\bf k}$ and energy transfer $\hbar\omega$ from a neutron.
Let ${\bf k}_{\rm i}^{}$ be the initial wave vector and
${\bf k}_{\rm f}^{}$ the final wave vector of the neutron,
 ${\bf k}={\bf k}_{\rm i}^{} - {\bf k}_{\rm f}^{}$
 and $\hbar\omega = \frac{\left|\hbar{\bf k}_{\rm i}^{}\right|^2-\left|\hbar{\bf k}_{\rm f}^{}\right|^2}{2 m}$,
 where $m$ is the mass of the neutron.
In the case of hydrogen atoms, the INS cross section is mainly due to incoherent contributions, and its dynamical structure factor is given by
\begin{equation}
 S_{\rm inc}^{}({\bf k},\omega) =
 \frac{1}{2\pi} \int_{-\infty}^{\infty} {\rm e}^{-{\rm i}\omega t}
 F_{\rm s}^{}({\bf k},t) {\rm d}t,
 \label{eq13}
\end{equation}
which is the Fourier transform of the self part (s) of the intermediate scattering function,
\begin{equation}
 F_{\rm s}^{}({\bf k},t) = \left\langle {\rm e}^{ -{\rm i}{\bf k}
 \cdot \hat{\bf R}_{\rm H}^{}(0) }
 {\rm e}^{ {\rm i}{\bf k}
 \cdot \hat{\bf R}_{\rm H}^{}(t) }
 \right\rangle,
 \label{eq14}
\end{equation}
where $\hat{\bf R}_{\rm H}^{}$ is the position operator of the hydrogen atom.
On the other hand, the Kubo-transformed type for the intermediate scattering function is \cite{miura1999path, craig2006inelastic}
\begin{equation}
 \tilde{F}_{\rm s}^{}({\bf k},t) = \frac{1}{\beta\hbar}
 \int_0^{\beta\hbar} {\rm d}\tau \left\langle {\rm e}^{ -{\rm i}{\bf k}
 \cdot \hat{\bf R}_{\rm H}^{}(-{\rm i}\tau) }
 {\rm e}^{ {\rm i}{\bf k}
 \cdot \hat{\bf R}_{\rm H}^{}(t) }
 \right\rangle,
 \label{eq15}
\end{equation}
and the associated dynamic structure factor is
\begin{equation}
 \tilde{S}_{\rm inc}^{}({\bf k},\omega) =
 \frac{1}{2\pi} \int_{-\infty}^{\infty} e^{-{\rm i}\omega t}
 \tilde{F}_{\rm s}^{}({\bf k},t) {\rm d}t.
 \label{eq16}
\end{equation}
Equations (\ref{eq13}) and (\ref{eq16}) are connected by the relationship
\begin{equation}
 S_{\rm inc}^{}({\bf k},\omega) =
 \frac{\beta\hbar\omega}{1-{\rm e}^{-\beta\hbar\omega}}
 \tilde{S}_{\rm inc}^{}({\bf k},\omega).
 \label{eq17}
\end{equation}
Following Eq. (\ref{eq10}), we assume
\begin{equation}
 \tilde{F}_{\rm s}^{\rm bcmd}({\bf k},t) =
 \left\langle  \overline{A}({\bf k},0)
 \overline{A}^\ast({\bf k},t)
 \right\rangle_{\rm bcmd}
 \label{eq18}
\end{equation}
as an approximation of Eq. (\ref{eq15}),
introducing the bead average
\begin{equation}
 \overline{A}({\bf k},t) = \frac{1}{P}
 \sum_{s=1}^P {\rm e}^{ -{\rm i}{\bf k}
 \cdot {\bf R}_{\rm H}^{(s)}(t) }.
 \label{eq19}
\end{equation}
Using Equations (\ref{eq16})--(\ref{eq18}),
we obtain
\begin{eqnarray}
 && S_{\rm inc}^{\rm bcmd}({\bf k},\omega)
 \nonumber \\
 &=& \frac{\beta\hbar\omega}{1-{\rm e}^{-\beta\hbar\omega}}
 \frac{1}{2\pi} \int_{-\infty}^{\infty}
 \left\langle \overline{A}({\bf k},0) \overline{A}^\ast({\bf k},t)
 \right\rangle_{\rm bcmd} e^{-{\rm i}\omega t} {\rm d}t.
 \nonumber \\
 \label{eq20}
\end{eqnarray}
Applying the Wiener-Khinchin theorem
 to Eq. (\ref{eq20}),
 we arrive at the final expression
 for the BCMD dynamic structure factor,
\begin{eqnarray}
 && S_{\rm inc}^{\rm bcmd}({\bf k},\omega)
 \nonumber \\
 &=& \frac{\beta\hbar\omega}{1-{\rm e}^{-\beta\hbar\omega}}
 \frac{1}{2\pi} \left\langle 
 \frac{1}{\tau} \left| \int_{-\frac{\tau}{2}}^{\frac{\tau}{2}}
 \overline{A}({\bf k},t) e^{-{\rm i}\omega t} {\rm d}t \right|^2 \right\rangle_{\rm bcmd},
 \nonumber \\
 \label{eq21}
\end{eqnarray}
which holds for a long trajectory length $\tau$.
Hereafter we call Eq. (\ref{eq21}) the calculated ``{\it INS spectrum}''.

{
In BCMD, the beadwise average is employed, so the canonical correlation function is calculated exactly at time at zero as in RPMD. This is in contrast with (adiabatic or partially adiabatic) CMD where the centroid average is employed.
Thus, Equations (\ref{eq19})--(\ref{eq21}) apply to RPMD as well,
and apply to MD with $P=1$.
In CMD, Eq. (\ref{eq19}) is changed to
\begin{eqnarray}
\overline{A}^{}({\bf k},t) = {\rm e}^{ -{\rm i}{\bf k}
 \cdot \overline{\bf R}_{\rm H}^{}(t) }
 \label{eq22}
\end{eqnarray}
 to be a function of the centroid coordinates,
 $\overline{\bf R}_{\rm H}^{}(t)=\frac{1}{P}\sum_{s=1}^P{\bf R}_{\rm H}^{(s)}(t)$.
}

\subsection{Vibrational density of states}

{
The vibrational density of states is defined by
\begin{eqnarray}
 D(\omega) = 4\pi \beta \sum_I^{N} M_I^{} \tilde{f}_I^{}(\omega)
 \label{eq23}
\end{eqnarray}
where the function
\begin{eqnarray}
 \tilde{f}_I^{}(\omega) = \frac{1}{2\pi} \int_{-\infty}^\infty dt e^{-{\rm i}\omega t} \tilde{c}_{I}^{}(t)
 \label{eq24}
\end{eqnarray}
is the Fourier transform of the canonical correlation function with respect to the velocity of atom $I$,
\begin{eqnarray}
 \tilde{c}_I^{}(t) = \frac{1}{\beta\hbar} \int_0^{\beta\hbar} d\tau \left\langle {\hat {\bf V}}_I^{}(t) {\hat {\bf V}}_I^{}(-{\rm i}\tau) \right\rangle.
 \label{eq25}
\end{eqnarray}
In Eq. (\ref{eq23}), the factor $4\pi\beta$ normalizes $D(\omega)$
to the vibrational degrees of freedom as $\int_0^{\infty} D(\omega) d\omega = 3N$
in the limit of harmonic oscillator systems
and high temperature/classical systems \cite{lin2003two}.
Following Eq. (\ref{eq10}), we assume 
\begin{eqnarray}
 \tilde{c}_I^{\rm bcmd}(t) = \left\langle \overline{\bf V}_I^{}(t) \overline{\bf V}_I^{}(0) \right\rangle_{\rm bcmd}^{}
 \label{eq26}
\end{eqnarray}
where
\begin{eqnarray}
 \overline{\bf V}_I^{}(t) = \frac{1}{P} \sum_{s=1}^P \dot{\bf R}_I^{(s)}(t).
 \label{eq27}
\end{eqnarray}
Applying the Wiener-Khinchin theorem, the final expression
 for the BCMD vibrational density of states is
\begin{eqnarray}
 D^{\rm bcmd}(\omega) =
 {2\beta}
 \sum_{I=1}^{N} M_I^{} \frac{1}{\tau}
 \left| \int_0^\tau \overline{\bf V}_I^{}(t)
 e^{-{\rm i}\omega t} {\rm d}t \right|^2.
 \label{eq28}
\end{eqnarray}
}


{
Equations (\ref{eq26})--(\ref{eq28}) apply to CMD and RPMD as well,
and apply to MD with $P=1$.
}

\section{Computational details}

For atomic interactions in the Pd-H system, we employed
 the Behler-Parrinello type ANN potential \cite{behler2007generalized,behler2015constructing,behler2021four} which has been modeled in our previous study \cite{kimizuka2022artificial}.
{Hydrogen is expected to exist as a neutral atom in Pd, and the ANN potential can mimic the DFT potential well.}
An ANN with Chebyshev descriptors for the radial and angular distribution functions was trained using the \texttt{\AE NET}) package \cite{artrith2016implementation,artrith2017efficient}.
The training set consisted of DFT calculations based on Perdew-Burke-Ernzerhof (PBE) exchange correlation functional \cite{perdew1996generalized} in the generalized gradient approximation (GGA), using \texttt{VASP})
\cite{kresse1996efficient}.
The root-mean-square errors of energy of the training and test datasets were 1.0 and 1.2 meV/atom, respectively.
For more details on the ANN modeling methodologies, see Refs. \cite{nagai2020self,kobayashi2021self,kimizuka2022artificial}.


It is known that vibrational frequency of hydrogen in 
Pd tends to be overestimated in the local density approximation of DFT when anharmonic contributions are considered \cite{elsasser1991vibrational,kimizuka2018mechanism}.
The present ANN potential was trained by the GGA with the PBE functional which corrects this overestimation.


A periodic system consisting of a cubic box containing 108 Pd atoms and 1 hydrogen atom was set up, where the Pd atoms were arranged in the fcc lattice.
The lattice constant was set to 3.942 \AA\ in the case of zero strain \cite{kimizuka2018mechanism}.
The BCMD simulations were performed for this system in canonical ensemble at temperatures from 50 to 300 K with the number of beads chosen to be 64.
At each temperature, 15--30 BCMD trajectories of length 5 ps with a step size of 0.5 fs were created.
They were restarted from different structures of thermal equilibrium obtained from preliminary PIMD simulations.
The INS spectrum, $S(k,\omega)$, was computed from the ensemble of BCMD trajectories according to Eq. (\ref{eq21}), and the ${\bf k}$ vectors were randomly sampled from all directions in three ranges of $k = |{\bf k}| =$ 0--4 \AA$^{-1}$, 4--8 \AA$^{-1}$, and 8--12 \AA$^{-1}$.
A hydrogen atom was placed either at the most stable O site, or the metastable tetrahedral (T) site, and the trajectories were sampled for those where the hydrogen atom stayed at the initial site.


For comparison, classical MD, CMD, and RPMD simulations were performed in a similar manner to the BCMD simulations.
For the CMD simulations, the adiabatic parameter was set to $\gamma_{\rm cmd}^{-1}=4$ and the step size was set to 0.1 and 0.005 fs for the centroid and non-centroid modes, respectively.
All calculations were performed using \texttt{PIMD} software \cite{shiga2001unified,shiga2023pimd},
with the implementation of hierarchical parallel computation with respect to beads and atomic interactions \cite{ruiz2016hierarchical}.


\section{Results}

\begin{figure}[htbp]
\includegraphics[width=0.9\columnwidth]{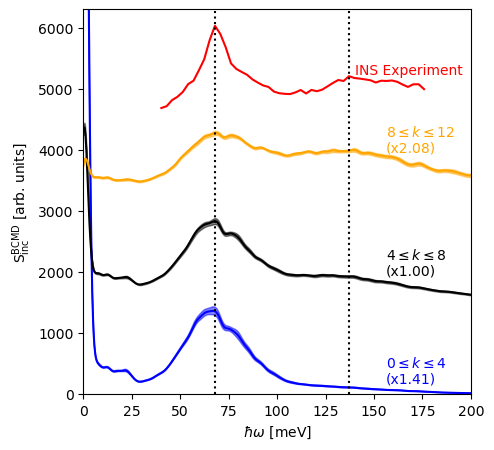}
\caption{
Calculated inelastic neutron scattering (INS) spectra by the semiclassical Brownian chain molecular dynamics (BCMD) simulations ($P=64$) at temperature 300 K\@ with no strain in the ranges $0\le k \le 4$ \AA$^{-1}$\ (blue), $4\le k \le 8$ \AA$^{-1}$\ (black), and $8\le k \le 12$ \AA$^{-1}$\ (orange), for the O site, and experimental INS spectrum at 295 K (red) \cite{rush1984direct}.
The data are displayed with the $y$ axis shifted by 1500 units each, and they are scaled by factors in parenthesis. The BCMD results are shown along with the range of statistical error in light color. The vertical dots indicate the two peak positions of the INS experiment.
}
\label{fig1}
\end{figure}

\begin{figure}[htbp]
\includegraphics[width=0.9\columnwidth]{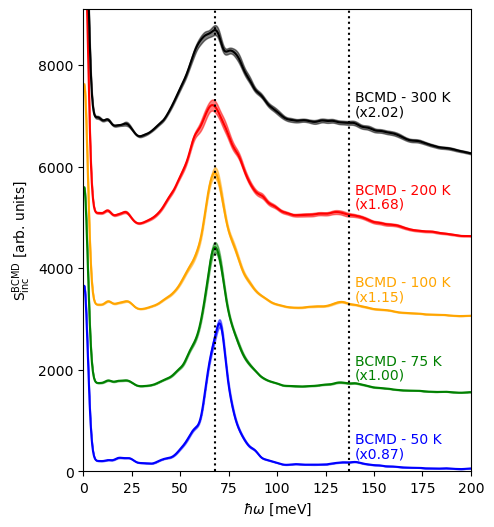}
\caption{Calculated inelastic neutron scattering (INS) spectra by the semiclassical Brownian chain molecular dynamics (BCMD) simulations ($P=64$) at temperatures 50 K (blue), 75 K (green), 100 K (orange), 200 K (red), and 300 K (black) for the O site in the range $4\le k \le 8$ \AA$^{-1}$\ with no strain. The data are displayed with the $y$ axis shifted by 1500 units each, and they are scaled by factors in parenthesis. The results are shown along with the range of statistical error in light color. The vertical dots indicate the two peak positions of the INS experiment at 295 K \cite{rush1984direct}.}
\label{fig2}
\end{figure}

\begin{figure}[htbp]
\includegraphics[width=0.9\columnwidth]{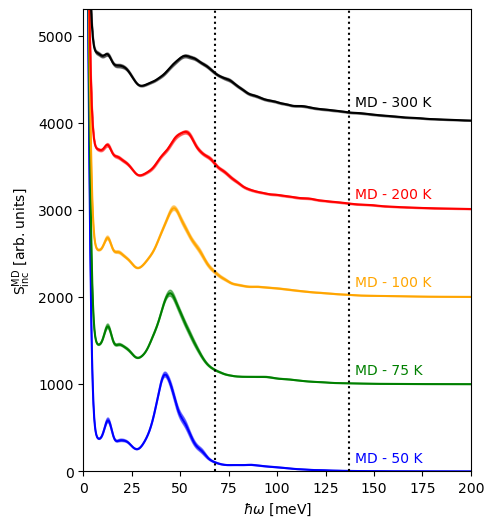}
\caption{Calculated inelastic neutron scattering (INS) spectra by the classical molecular dynamics (MD) simulations ($P=1$) at temperatures 50 K (blue), 75 K (green), 100 K (orange), 200 K (red), and 300 K (black) for the O site in the range $8\le k \le 12$ \AA$^{-1}$\ with no strain. The data are displayed with the $y$ axis shifted by 1000 units each. The results are shown along with the range of statistical error in light color. The vertical dots indicate the two peak positions of the INS experiment at 295 K \cite{rush1984direct}.}
\label{fig3}
\end{figure}

\begin{figure}[htbp]
\includegraphics[width=0.7\columnwidth]{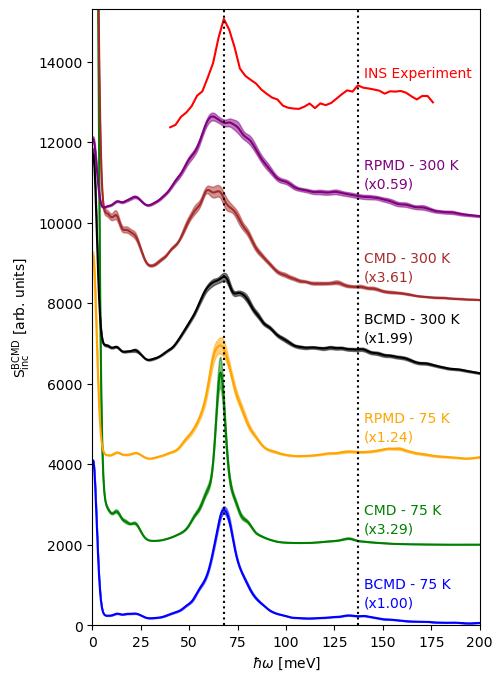}
\caption{
Calculated inelastic neutron scattering (INS) spectra by the semiclassical Brownian chain molecular dynamics (BCMD) simulations ($P=64$) at temperatures 75 K (blue) and 300 K (black), the semiclassical centroid molecular dynamics (CMD) simulations ($P=64$) at temperatures 75 K (green) and 300 K (brown), and the semiclassical ring polymer molecular dynamics (RPMD) simulations ($P=64$) at temperatures 75 K (orange) and 300 K (purple), for the O site in the range $4\le k \le 8$ \AA$^{-1}$\ with no strain, and experimental INS spectrum at 295 K (red) \cite{rush1984direct}. The data are displayed with the $y$ axis shifted by 2000 units each, and they are scaled by factors in parenthesis. The results are shown along with the range of statistical error in light color. The vertical dots indicate the two peak positions of the INS experiment.}
\label{fig4}
\end{figure}

\begin{figure}[htbp]
\includegraphics[width=0.9\columnwidth]{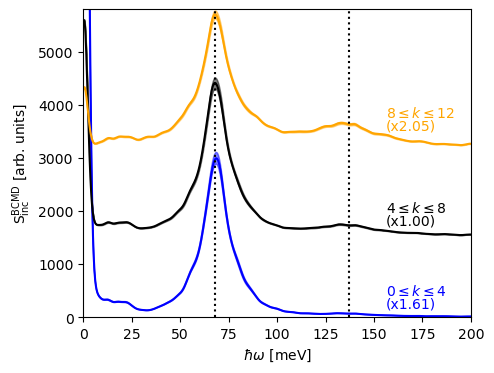}
\caption{
Calculated inelastic neutron scattering (INS) spectra by the semiclassical Brownian chain molecular dynamics (BCMD) simulations ($P=64$) at temperature 75 K with no strain in the ranges $0\le k \le 4$ \AA$^{-1}$\ (blue), $4\le k \le 8$ \AA$^{-1}$\ (black), and $8\le k \le 12$ \AA$^{-1}$\ (orange), for the O site.
The data are displayed with the $y$ axis shifted by 1500 units each, and they are scaled by factors in parenthesis. The results are shown along with the range of statistical error in light color. The vertical dots indicate the two peak positions of the INS experiment at 295 K \cite{rush1984direct}.
}
\label{fig5}
\end{figure}

\begin{figure}[htbp]
\includegraphics[width=0.9\columnwidth]{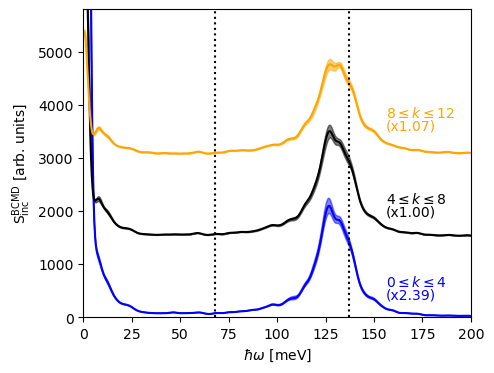}
\caption{
Same as Fig. \ref{fig5}, but for the the T site.
}
\label{fig6}
\end{figure}

\begin{figure}[htbp]
\includegraphics[width=0.9\columnwidth]{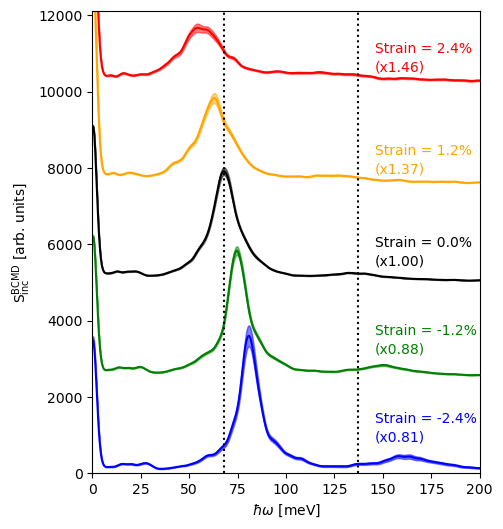}
\caption{
Calculated inelastic neutron scattering (INS) spectra by the semiclassical Brownian chain molecular dynamics (BCMD) simulations ($P=64$) at temperature 75 K\@ for the O site in the range $4\le k \le 8$ \AA$^{-1}$\ with hydrostatic (axial) strains of -2.4\% (blue), -1.2\% (green), 0\% (black), 1.2\% (orange), and 2.4\% (red).
The data are displayed with the $y$ axis shifted by 2500 units each, and they are scaled by factors in parenthesis. The results are shown along with the range of statistical error in light color. The vertical dots indicate the two peak positions of the INS experiment at 275 K \cite{rush1984direct}.
}
\label{fig7}
\end{figure}

The results of the calculated INS spectra are displayed in Figs.
\ref{fig1}--\ref{fig7},
and the peak positions are listed in Table \ref{tab1}.
When the peak positions are not clear, they are obtained by fitting the
 spectra to the sum of two Lorenzian functions.



%
Figure \ref{fig1} compares the INS spectra calculated from the semiclassical BCMD simulation at 300 K for the O site with that obtained from the INS experiment of PdH$_{0.014}^{}$ at 295 K \cite{rush1984direct}.
Since the $k$ value was not reported in the experimental paper, the calculated results were displayed for different ranges of $k=|{\bf k}|$.
The BCMD spectrum consists of a sharp and tall first peak $\sim$68 meV, and a broad and short second peak $\sim$139 meV, which captures the characteristics of the experimental spectrum with the peak positions of $69.0 \pm 0.5$ and $137 \pm 2$ meV\@.
These peaks represent the fundamental tones and overtones, respectively, of the vibrational excitation of the hydrogen atom.
The intensity of the shoulder of the first peak at $\sim$80 meV increases with the temperature as in Fig. \ref{fig2},
so it is presumed to be a sideband due to coupling with the Pd phonon.
This should correspond to the asymmetry of the first peak in the experimental spectrum.


The calculated spectral width at 300 K cannot be directly compared with the experimental one at 295 K, because it changes significantly with the $k$ range. 
However, even if choosing $4 \le k \le 8$ \AA$^{-1}$\ that has a resemblance, the BCMD spectral width looks broader than the experimental one.
The same trend is seen not only in the results of other semiclassical methods, CMD and RPMD, but also in the results of classical MD {where the broadening is expected to have been more limited in the absence of NQEs}, see Figs. \ref{fig3} and \ref{fig4}.
Thus, it is likely that the discrepancy with experiment on the spectral width is not due to the semiclassical approximation, but rather the PBE functional that the ANN PES is based on.
{
This point is further discussed in the next section.
}


Figure \ref{fig2} shows that the peak position of the BCMD spectrum changes little with temperature.
The peaks become visibly less intense and more broad as the temperature is increased from 100 to 300 K\@, while they change little $\leq$100 K\@.
The intensity of the second peak relative to that of the first peak
 increases with temperature,
 because higher energy vibrational states contributes
 to the correlation function in Eq. (\ref{eq18}).
These trends in the temperature dependence of the peak intensity
 are like those measured in a recent INS experiment of
 metal hydride, ZrV$_2^{}$H$_x^{}$ \cite{borgschulte2020inelastic},
 although the peak positions are different from those of PdH$_x^{}$.


%
The semiclassical BCMD spectrum in Fig. \ref{fig2} is significantly blueshifted compared with the classical MD spectrum shown in Fig. \ref{fig3}.
This indicates that NQE is present in the hydrogen vibrations and essential to reproduce the experimental spectrum.
The blue-shift is attributed to the combination of the NQE of hydrogen atoms and the anharmonic potential with even symmetry.
The NQE of hydrogen atoms appears as large amplitude zero-point vibration on the anharmonic  PES\@.
Since the PES has an even symmetry reflecting repulsive interactions with Pd atoms on both sides, the leading order of anharmonicity is quartic.
For this reason the NQE acts as a blue-shift in the vibrational spectra in this system.
This is in contrast with the fact that the NQE of the hydrogen atom in molecular systems (OH, CH bonds, etc.) usually acts as a red-shift where the leading order of anharmonicity is cubic.


For classical MD at zero temperature, the frequency of the first peak is expected to approach the harmonic frequency (HAR), which is significantly underestimated to be 34.8 meV by DFT calculations based on the PBE functional.
As NQEs are weakened with increasing temperature $T$, the classical MD and BCMD spectra should approach each other.
For this reason the first peak of classical MD is blue-shifted with increasing temperature.
Even at $T = 300$ K, the spectral difference between classical MD and BCMD is still very large, which means that NQEs are important at room temperature.
In addition, the intensity of the second peak of classical MD is much weaker than that of BCMD\@.
{As discussed in the next section, this could be understood from the weaker vibrational coupling to the Pd phonon in the absence of hydrogen zero-point vibrations.}
%


Figure \ref{fig4} compares the calculated spectra of CMD and RPMD at 75 and 300 K\@ with the experimental ones at 295 K\@.
The first peaks of CMD and RPMD spectra do not differ significantly from the experimental one, while the second peaks of the CMD and RPMD spectra are slightly red-shifted and slightly split, respectively, from the experimental one. 
These shortcomings of CMD and RPMD are known as curvature and chain resonance problems, respectively, which become pronounced at low temperature.
The former arises from a spurious coupling between rotations and vibrations due to the adiabatic separation of the centroid and noncentroid variables, while the latter is due to spurious resonance between the vibrations of the physical mode and the cyclic chain \cite{witt2009applicability,ivanov2010artificial}.
Because BCMD reduces these problems, the BCMD spectra are in better agreement with the experimental one than the CMD and RPMD spectra with respect to the peak positions.


Figure \ref{fig5} shows the $k$ dependence of the calculated INS spectra at 75 K\@. The trend is like those shown in Fig. \ref{fig1}, except that the calculated INS spectra at 75 K are consistently sharper in their shape than the ones at 300 K\@. As $k$ increases, the spectra become slightly broader, but the peak positions remain almost the same.
{
The $k$-dependence arises from the state-to-state scattering intensity factor, $I_{mn}^{}({\bf k})$, as can be seen
 from the wavefunction representation for Eq. (\ref{eq13}) as
\begin{eqnarray}
 S_{\rm inc}^{}({\bf k},\omega) =
 \frac{ \sum_{m,n} \delta(\omega-\omega_m^{}+\omega_n^{}) e^{-\beta\hbar\omega_n^{}}
 I_{mn}^{}({\bf k}) }
 {\sum_n e^{-\beta\hbar\omega_n^{}}},
 \label{eq29}
\end{eqnarray}
 where $\psi_m^{}$ and $\omega_m^{}$ denote the normalized eigenfunction and eigenfrequency, respectively, of the $m$-th state
 of the coupled hydrogen-Pd system, and
\begin{eqnarray}
 I_{mn}^{}({\bf k}) = \left|\left\langle \psi_m^{} \right| {\rm e}^{ {\rm i}{\bf k}
 \cdot \hat{\bf R}_{\rm H}^{} } \left| \psi_n^{}  \right\rangle \right|^2.
 \label{eq30}
\end{eqnarray}
}
{
Note that the subscripts $m$ and $n$ refer to the eigenstates of the entire system, not the hydrogen vibration alone. Contributions from the Pd phonon side band at the frequency $\omega=\omega_m^{}-\omega_n^{}$ is dependent on ${\bf k}$ by its weight $I_{mn}^{}({\bf k})$.
}


The calculated INS spectra of hydrogen atoms in the T site have not been measured, but it is believed that the T site could be occupied in Pd nanocrystals with lattice distortion or Pd surface in a non-equilibrium environment \cite{akiba2016nanometer}. 
Thus it may be of future interest to predict the INS spectra for the T site. 
Figure \ref{fig6} shows the calculated spectra for the T site under the same conditions as in Fig. \ref{fig5}. 
Two peaks of the INS spectra appear at 127 and 136 meV\@. The latter is close to the HAR frequency of 135 meV\@. 
The peak splitting is presumably due to the coupling of hydrogen vibration with the Pd phonon.

\begin{figure}[htbp]
\includegraphics[width=0.9\columnwidth]{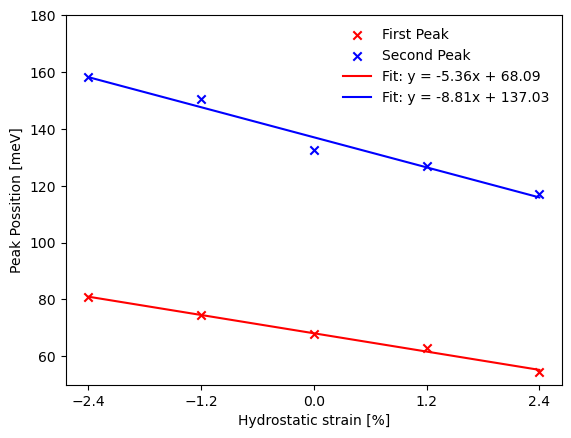}
\caption{
The plots of the peak positions of Fig. \ref{fig7} with respect to hydrostatic (axial) strain. The fit to linear function is shown as a guide to the eye.
}
\label{fig8}
\end{figure}

Lattice distortions on the order of a few percent are often observed in local regions of Pd nanostructures containing concentrated hydrogen solid solutions, hydrides, defects, impurities, heterophase boundaries, etc.
Since change in the INS spectra is detected associated with lattice distortion, it is important to provide its theoretical foundation.
Here we studied a model case of hydrogen in Pd under hydrostatic (axial) strains $-2.4{\rm \%} \le \epsilon \le 2.4 {\rm \%}$ (where positive $\epsilon$ means expansive in this definition).
Figure \ref{fig7} shows that the first peak of INS spectrum is monotonically red-shifted with increasing strain from negative to positive.
As the Pd lattice expands, the repulsive force with the Pd atoms decreases and the curvature of the hydrogen potential decreases, resulting in a decrease in the vibrational frequency of the hydrogen atoms.
As shown in Fig. \ref{fig8}, the first and second peak positions can be fitted to a linear function,
\begin{eqnarray}
 \hbar\omega_1^{} &\approx& 68.1 - 536 \epsilon\ ({\rm meV}) \ {\rm and}
 \nonumber \\ 
 \hbar\omega_2^{} &\approx& 137.0 - 881 \epsilon \ ({\rm meV}),
 \label{eq31}
\end{eqnarray}
respectively.
The red-shift upon positive lattice strain is consistent with recent INS measurements of  nanocrystalline PdH$_{0.42}^{}$ where the first peak is found at 59.3 eV in expanded Pd lattice of a few percent \cite{akiba2016nanometer,kofu2016hydrogen,kofu2017vibrational,kofu2020dynamics}.

\section{Discussion}

\begin{figure}[htbp]
{
\includegraphics[width=0.4\columnwidth]{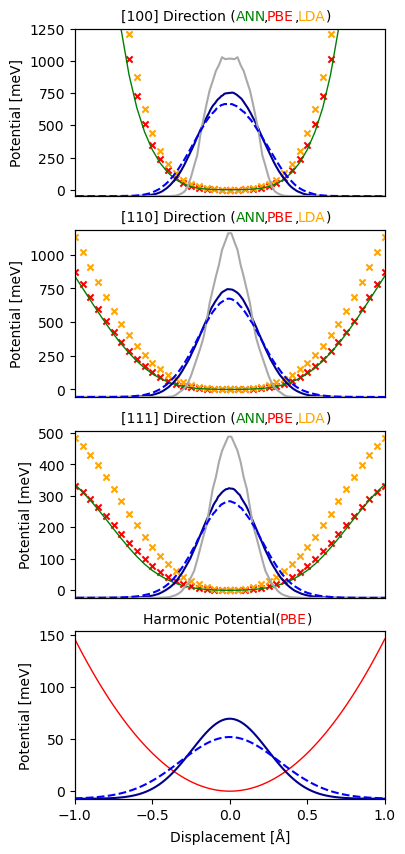}
\caption{Top three panels show the plots of potential energy function as a function of hydrogen atom displacement from the minimum of the O site along the directions [100], [110], and [111],
for the artificial neural network (ANN) potential used in the study (green line), the density functional theory (DFT) potential based on
the PBE functional (red crosses), and the local density approximation (LDA) functional (orange crosses). 
Here we also display the hydrogen distributions at temperatures 75 K
 (solid dark blue line) and 300 K (broken blue line) obtained
 from the quantum path integral molecular dynamics (PIMD) simulations,
 and at temperatures 75 K
 obtained from the classical molecular dynamics (MD) simulations (solid gray line), in the respective
 directions using the ANN potential.
The bottom panel shows the harmonic potential function with a curvature at the minimum of the O site in a given direction for the DFT potential based on the PBE functional (red), with an eigenfrequency $\omega_{\rm h}^{} =$ 34.8 meV (triply degenerate). We also display the hydrogen distributions at temperatures 75 K (solid dark blue line) and 300 K (broken blue line) that would be obtained in this harmonic potential, i.e., $\rho(x)= \sqrt{\frac{\alpha}{\pi}} \exp\left(-\alpha x^2\right)$ with $\alpha=\frac{m\omega}{\hbar} \tanh\left(\frac{\beta\hbar\omega}{2}\right)$.}
\label{fig9}
}
\end{figure}

{
The following analysis was performed to comprehend the results of the calculated INS spectra.
Figure \ref{fig9} shows the DFT and ANN potential energy curves calculated along the [100], [110], and [111] directions.
It is important to note that these potential curves are strongly anharmonic for all directions of hydrogen vibrations.
The quantum distribution of hydrogen on the anharmonic potential is considerably narrower than the quantum distribution on the harmonic potential.
This is consistent with our result that the anharmonicity acts as a blueshift in the vibrational spectra.
}

\begin{figure}[htbp]
{
\includegraphics[width=1.0\columnwidth]{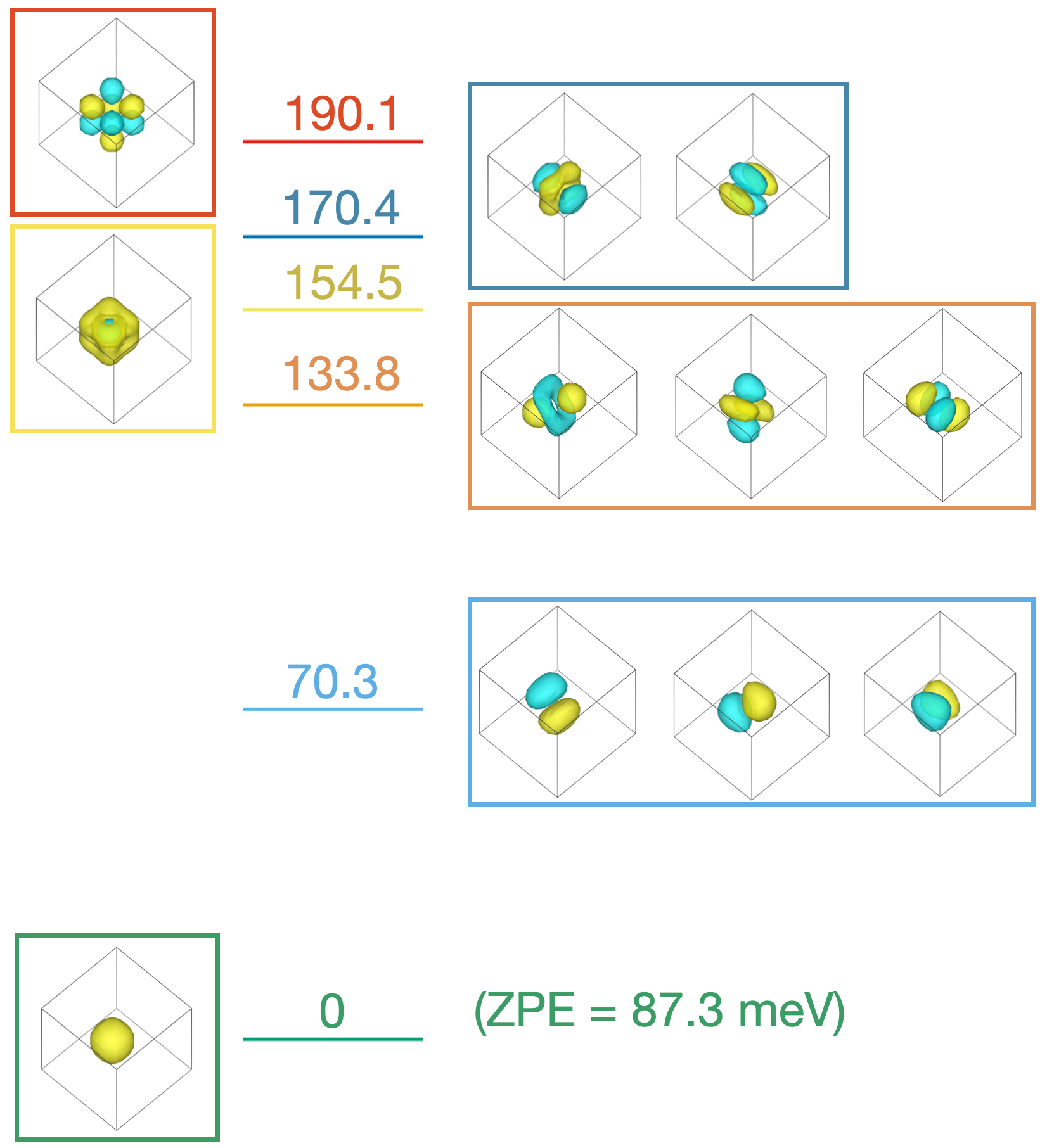}
\caption{Adiabatic vibrational energy levels of hydrogen atom at the O site using the artificial neural network (ANN) potential energy surface. Here the coupling to Pd phonons is neglected. The energies relative to the ground state are shown in millielectronvolts, and the corresponding hydrogen wavefunctions are depicted. The zero point energy (ZPE) of the ground state is estimated as 87.3 meV.}
\label{fig10}
}
\end{figure}

{
As a different approach, adiabatic vibrational energy levels of the hydrogen atom in the O site were calculated by solving the three-dimensional time-independent Schr\"odinger equation. The PES was calculated by the ANN potential as a function of the hydrogen atom displacement while fixing the Pd atom in an optimized geometry. This is a kind of adiabatic approximation, in which the anharmonicity of hydrogen vibration is considered, while the coupling to the Pd phonons is neglected. The discrete variable representation (DVR) technique \cite{colbert1992novel} was used for the PES described by $17\times 17 \times 17$ regular grids placed at $\pm 1$ \AA\ around the O-site minimum.
As shown in Fig. \ref{fig10}, the fundamental tone, 70.3 meV, and the first overtone, 133.8 meV, agree well with the INS peak positions obtained from the BCMD simulations as displayed in Table \ref{tab1}.
This result ensures the role of anharmonicity in controlling the peak position of the vibrational spectra, and the reliability of the PBE functional reflected in the ANN PES to correctly estimate the peak position.
}


{
As Fig. \ref{fig10} shows, both the excited states responsible for the fundamental tone and first overtone are triply degenerate because of the spatial symmetry of the O site, in the absence of Pd phonon coupling. This means that this approach cannot account for the side bands in the INS spectra. Phonon coupling, which is missing in this approach, is responsible for the side bands in the INS spectra. For this reason, the origin of the shoulder observed $\sim$80 meV is expected to be the hydrogen-Pd coupling.
We note that Figure \ref{fig10} is basically consistent with a recent experimental and computational study by Ozawa {\it et al.} \cite{ozawa2023observation}.
}

\begin{figure}[htbp]
{
\includegraphics[width=0.6\columnwidth]{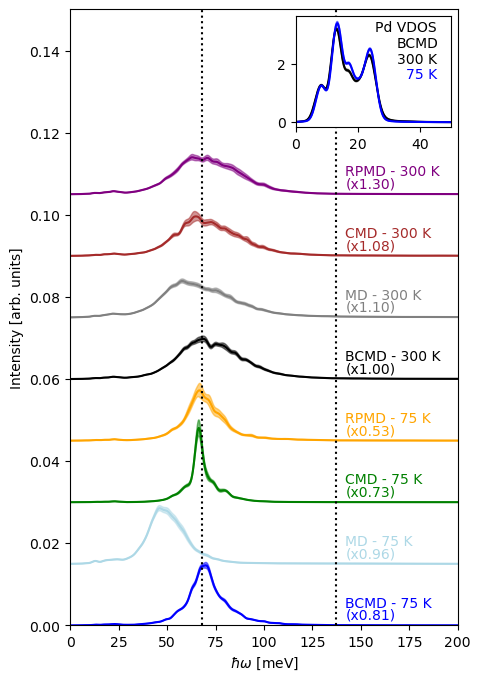}
\caption{Hydrogen contribution to vibrational density of states calculated by the semiclassical Brownian chain molecular dynamics (BCMD) simulations ($P=64$) at temperatures 75 K (blue) and 300 K (black), the classical molecular dynamics (MD) simulations ($P=1$) at at temperatures 75 K (cyan) and 300 K (gray), the semiclassical centroid molecular dynamics (CMD) simulations ($P=64$) at temperatures 75 K (green) and 300 K (brown), and the semiclassical ring polymer molecular dynamics (RPMD) simulations ($P=64$) at temperatures 75 K (orange) and 300 K (purple), for the O site in the range $4\le k \le 8$ \AA$^{-1}$\ with no strain, and experimental inelastic neutron scattering (INS) spectrum at 295 K (red) \cite{rush1984direct}. The data are displayed with the y-axis shifted by 0.01 units each, and they are scaled by factors in parenthesis. The results are shown along with the range of statistical error in light color. The vertical dots indicate the two peak positions of the INS experiment. Inset panel shows the Pd contribution to vibrational density of states calculated by the semiclassical BCMD simulations ($P=64$) at temperatures 75 K (blue) and 300 K (black).}
\label{fig11}
}
\end{figure}

{
The vibrational density of states, which represents the set of single-phonon vibrational frequencies,
 is shown in Fig. \ref{fig11}.
The vibrational density of states covers the region of the first peak in the INS spectra
 representing the fundamental tone.
As expected, it does not cover the second peak in the INS spectra representing the overtones containing
 multiple phonons.
Interestingly, the vibrational density of states
 covers most of the side bands of the INS spectra for the fundamental tone,
 suggesting that the side bands consists of a mixture of hydrogen and Pd vibrations.
}

{
The reason the intensity of the second peak in the INS spectra is weaker than the first peak is that overtone excitation is generally more difficult than fundamental excitation. Overtone excitation of hydrogen requires a large transition matrix via strong coupling to the Pd phonon which is expected to be amplified by the magnitude of the hydrogen vibration. In the presence of zero-point vibrations, the NQE increases the magnitude of hydrogen vibration, as can be seen in Fig. \ref{fig9}. This causes the intensity of the second peak is weaker in the classical MD simulations than in the semiclassical BCMD, CMD, and RPMD simulations.
}


{
Figure \ref{fig9} confirms that the ANN potential reproduce well the DFT potential based on the PBE functional.
However, as can be inferred from the difference between the LDA and PBE functions shown in Fig. \ref{fig9}, the functional dependence on the DFT potentials may have a non-negligible impact on the spectral line shape.
There might be room for improvement using a DFT functional that is more accurate than GGA, but we leave this as an issue for future research.
}


%
\section{Conclusions}

The semiclassical BCMD is a general computational approach that consistently incorporates NQE and anharmonic effects of vibration properties at finite temperatures. Combined with ANN potentials of DFT-level accuracy, it provides a reliable prediction of vibrational spectra for condensed matter systems.
In this paper, the method was found to be effective in calculating INS spectra of hydrogen in metal Pd. With NQEs considered, it accurately calculates the peak positions of the spectra and qualitatively reproduces the spectra in terms of shape. Along this line, computational prediction of INS spectra of hydrogen atoms trapped at metastable sites and in heterogeneous environments, etc., is expected to be useful in understanding the spectra measured at various experimental conditions.

\begin{acknowledgments}

We thank the JSPS Grant-in-Aid for Scientific Research (Grants No. 23K04670, No. 21H01603, 23H01273, and No. 18H05519) for financial support.
The calculations were conducted using the supercomputer facilities at Japan Atomic Energy Agency.
We thank Prof. Shinichi Miura in Kanazawa University for his advice on coding the dynamic structure factor, and Dr. Maiko Kofu for discussion on INS experiments.
\end{acknowledgments}

\appendix

\section{Bead convergence}

\begin{figure}[htbp]
\includegraphics[width=0.9\columnwidth]{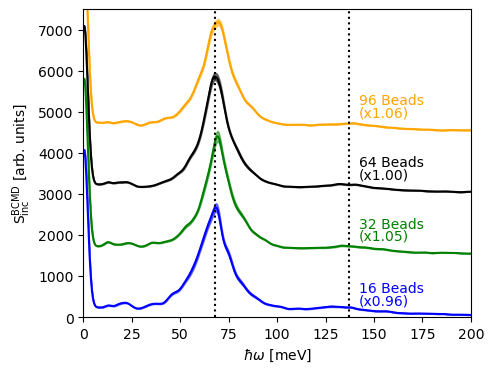}
\caption{Calculated inelastic neutron scattering (INS) spectra by the Brownian chain molecular dynamics (BCMD) simulations at temperature 75 K using $P=16$ (blue), $P=32$ (green), $P=64$ (black), and $P=96$ (orange) for the O site in the range $4\le k \le 8$ \AA$^{-1}$\ with no strain. The data are displayed with the $y$ axis shifted by 1000 units each, and they are scaled by factors in parenthesis.
The BCMD results are shown along with the range of statistical error in light color. The vertical dots indicate the two peak positions of the INS experiment at 295 K \cite{rush1984direct}.}
\label{fig12}
\end{figure}

As Fig. \ref{fig12} shows, the INS spectra from the semiclassical BCMD simulations were not significantly different from each other when the number of beads was set as $P \ge 64$.
Therefore, the main results of this paper are presented for the $P=64$ case.

\section{Imaginary time intermediate scattering function}

{
Following Eq. (\ref{eq14}), the evolution in imaginary time
$\tau={\rm i}t$ of the intermediate scattering function is
expressed as
\begin{eqnarray}
 F_{\rm s}^{}\left({\bf k},-{\rm i}\tau\right)
 = \left\langle {\rm e}^{-{\rm i}{\bf k}\cdot\hat{\bf R}_{\rm H}^{}(0)}
 {\rm e}^{{\rm i}{\bf k}\cdot\hat{\bf R}_{\rm H}^{}(-{\rm i}\tau)}
 \right\rangle.
 \label{eq:B1}
\end{eqnarray}
Eq. (\ref{eq:B1}) can be calculated rigorously by quantum PIMD simulations as
\begin{eqnarray}
 F_{\rm s}^{\rm pimd}(k,\tau) = \left\langle {\rm e}^{-{\rm i}{\bf k} \cdot {\bf R}_{\rm H}^{(s)}}
 {\rm e}^{{\rm i}{\bf k}\cdot {\bf R}_{\rm H}^{(u)}}  \right\rangle_{\rm pimd}^{},
 \label{eq:B2}
\end{eqnarray}
for a pair of beads $s$ and $u$ with $s-u = \frac{P\tau}{\beta\hbar}$.
On the other hand, the imaginary time intermediate scattering function can also be calculated from the real time information of $S_{\rm inc}^{}(\omega)$ using the Wick rotation \cite{perez2009comparative},
\begin{eqnarray}
 F_{\rm s}^{}\left({\bf k},-{\rm i}\tau\right)
 = \int_{-\infty}^\infty {\rm d}\omega
 S_{\rm inc}^{}(\omega)
 e^{-\frac{\beta\hbar\omega}{2}}
 \cosh\left( \frac{\beta\hbar\omega}{2} -\omega\tau\right).
\nonumber \\
 \label{eq:B3}
\end{eqnarray}
The proof of Eq. (\ref{eq:B3}) can be done by expanding the averages on both sides in terms of the eigenstates of the system Hamiltonian.
Eq. (\ref{eq:B3}) can be calculated approximately by $S_{\rm inc}^{}(\omega)$ obtained from classical MD simulations and semiclassical BCMD, CMD and RPMD simulations as $F_{\rm s}^{\rm md}(k,\tau)$, $F_{\rm s}^{\rm bcmd}(k,\tau)$, $F_{\rm s}^{\rm cmd}(k,\tau)$, and $F_{\rm s}^{\rm rpmd}(k,\tau)$, respectively.
The quality of the approximations of $S_{\rm inc}^{}(\omega)$ in the respective methods can thus be tested.
}

\begin{figure}[htbp]
{
\includegraphics[width=0.7\columnwidth]{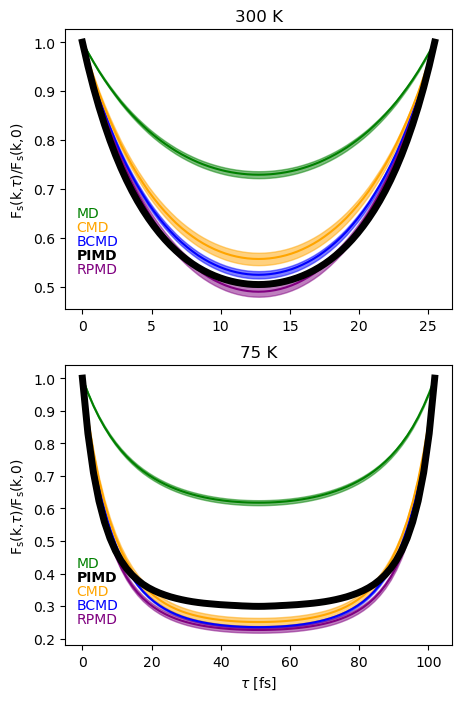}
\caption{Imaginary time intermediate scattering function $F_{\rm s}^{}(k,\tau)$ with $0 \le \tau \le \beta\hbar$ normalized by $F_{\rm s}^{}(k,0)$
 obtained from the classical molecular dynamics (MD; green), the semiclassical Brownian chain molecular dynamics (BCMD; blue),
the semiclassical centroid molecular dynamics (CMD; orange), and the semiclassical ring polymer molecular dynamics (RPMD; purple) simulations
at temperature 300 K (top panel) and 75 K (bottom panel) using Eq. (\ref{eq:B3}),
for the O site in the range $4\le k \le 8$ \AA$^{-1}$\ with no strain.
The results are shown along with the range of statistical error in light color.
The functions obtained from the quantum path integral molecular dynamics (PIMD) simulations (black)
using Eq. (\ref{eq:B2}) are also shown.}
\label{fig13}
}
\end{figure}

{
The results in Figure \ref{fig13} show that $F_{\rm s}^{\rm bcmd}(k,\tau)$, $F_{\rm s}^{\rm cmd}(k,\tau)$, and $F_{\rm s}^{\rm rpmd}(k,\tau)$ agree with $F_{\rm s}^{\rm pimd}(k,\tau)$ much better than $F_{\rm s}^{\rm md}(k,\tau)$ especially near both ends, $\tau \approx 0$ and $\beta\hbar$.
This indicates that these semiclassical approximations properly account for the NQEs.
However, as the imaginary time $\tau$ approaches the center of the thermal interval ($\tau=\beta\hbar/2$), the agreement with $F_{\rm s}^{\rm pimd}(k,\tau)$ deteriorates for all   $F_{\rm s}^{\rm bcmd}(k,\tau)$, $F_{\rm s}^{\rm cmd}(k,\tau)$, and $F_{\rm s}^{\rm rpmd}(k,\tau)$.
}

\bibliography{skw.bib}

\begin{thebibliography}{84}%
\makeatletter
\providecommand \@ifxundefined [1]{%
 \@ifx{#1\undefined}
}%
\providecommand \@ifnum [1]{%
 \ifnum #1\expandafter \@firstoftwo
 \else \expandafter \@secondoftwo
 \fi
}%
\providecommand \@ifx [1]{%
 \ifx #1\expandafter \@firstoftwo
 \else \expandafter \@secondoftwo
 \fi
}%
\providecommand \natexlab [1]{#1}%
\providecommand \enquote  [1]{``#1''}%
\providecommand \bibnamefont  [1]{#1}%
\providecommand \bibfnamefont [1]{#1}%
\providecommand \citenamefont [1]{#1}%
\providecommand \href@noop [0]{\@secondoftwo}%
\providecommand \href [0]{\begingroup \@sanitize@url \@href}%
\providecommand \@href[1]{\@@startlink{#1}\@@href}%
\providecommand \@@href[1]{\endgroup#1\@@endlink}%
\providecommand \@sanitize@url [0]{\catcode `\\12\catcode `\$12\catcode
  `\&12\catcode `\#12\catcode `\^12\catcode `\_12\catcode `\%12\relax}%
\providecommand \@@startlink[1]{}%
\providecommand \@@endlink[0]{}%
\providecommand \url  [0]{\begingroup\@sanitize@url \@url }%
\providecommand \@url [1]{\endgroup\@href {#1}{\urlprefix }}%
\providecommand \urlprefix  [0]{URL }%
\providecommand \Eprint [0]{\href }%
\providecommand \doibase [0]{http://dx.doi.org/}%
\providecommand \selectlanguage [0]{\@gobble}%
\providecommand \bibinfo  [0]{\@secondoftwo}%
\providecommand \bibfield  [0]{\@secondoftwo}%
\providecommand \translation [1]{[#1]}%
\providecommand \BibitemOpen [0]{}%
\providecommand \bibitemStop [0]{}%
\providecommand \bibitemNoStop [0]{.\EOS\space}%
\providecommand \EOS [0]{\spacefactor3000\relax}%
\providecommand \BibitemShut  [1]{\csname bibitem#1\endcsname}%
\let\auto@bib@innerbib\@empty
\bibitem [{\citenamefont {Alefeld}\ and\ \citenamefont
  {V{\"o}lkl}(1978)}]{alefeld1978hydrogen}%
  \BibitemOpen
  \bibfield  {author} {\bibinfo {author} {\bibfnamefont {G.}~\bibnamefont
  {Alefeld}}\ and\ \bibinfo {author} {\bibfnamefont {J.}~\bibnamefont
  {V{\"o}lkl}},\ }\href@noop {} {\emph {\bibinfo {title} {Hydrogen in metals
  {I}-Basic properties}}}\ (\bibinfo  {publisher} {Springer-Verlag},\ \bibinfo
  {year} {1978})\BibitemShut {NoStop}%
\bibitem [{\citenamefont {Fukai}(2006)}]{fukai2006metal}%
  \BibitemOpen
  \bibfield  {author} {\bibinfo {author} {\bibfnamefont {Y.}~\bibnamefont
  {Fukai}},\ }\href@noop {} {\emph {\bibinfo {title} {The metal-hydrogen
  system: basic bulk properties}}}\ (\bibinfo  {publisher} {Springer Science \&
  Business Media},\ \bibinfo {year} {2006})\BibitemShut {NoStop}%
\bibitem [{\citenamefont {Adams}\ and\ \citenamefont
  {Chen}(2011)}]{adams2011role}%
  \BibitemOpen
  \bibfield  {author} {\bibinfo {author} {\bibfnamefont {B.~D.}\ \bibnamefont
  {Adams}}\ and\ \bibinfo {author} {\bibfnamefont {A.}~\bibnamefont {Chen}},\
  }\href@noop {} {\bibfield  {journal} {\bibinfo  {journal} {Mater. {T}oday}\
  }\textbf {\bibinfo {volume} {14}},\ \bibinfo {pages} {282} (\bibinfo {year}
  {2011})}\BibitemShut {NoStop}%
\bibitem [{\citenamefont {Bergsma}\ and\ \citenamefont
  {Goedkoop}(1960)}]{bergsma1960thermal}%
  \BibitemOpen
  \bibfield  {author} {\bibinfo {author} {\bibfnamefont {J.}~\bibnamefont
  {Bergsma}}\ and\ \bibinfo {author} {\bibfnamefont {J.}~\bibnamefont
  {Goedkoop}},\ }\href@noop {} {\bibfield  {journal} {\bibinfo  {journal}
  {Physica}\ }\textbf {\bibinfo {volume} {26}},\ \bibinfo {pages} {744}
  (\bibinfo {year} {1960})}\BibitemShut {NoStop}%
\bibitem [{\citenamefont {Chowdhury}\ and\ \citenamefont
  {Ross}(1973)}]{chowdhury1973neutron}%
  \BibitemOpen
  \bibfield  {author} {\bibinfo {author} {\bibfnamefont {M.}~\bibnamefont
  {Chowdhury}}\ and\ \bibinfo {author} {\bibfnamefont {D.}~\bibnamefont
  {Ross}},\ }\href@noop {} {\bibfield  {journal} {\bibinfo  {journal} {Solid
  State Commun.}\ }\textbf {\bibinfo {volume} {13}},\ \bibinfo {pages} {229}
  (\bibinfo {year} {1973})}\BibitemShut {NoStop}%
\bibitem [{\citenamefont {Drexel}\ \emph {et~al.}(1976)\citenamefont {Drexel},
  \citenamefont {Murani}, \citenamefont {Tocchetti}, \citenamefont {Kley},
  \citenamefont {Sosnowska},\ and\ \citenamefont {Ross}}]{drexel1976motions}%
  \BibitemOpen
  \bibfield  {author} {\bibinfo {author} {\bibfnamefont {W.}~\bibnamefont
  {Drexel}}, \bibinfo {author} {\bibfnamefont {A.}~\bibnamefont {Murani}},
  \bibinfo {author} {\bibfnamefont {D.}~\bibnamefont {Tocchetti}}, \bibinfo
  {author} {\bibfnamefont {W.}~\bibnamefont {Kley}}, \bibinfo {author}
  {\bibfnamefont {I.}~\bibnamefont {Sosnowska}}, \ and\ \bibinfo {author}
  {\bibfnamefont {D.}~\bibnamefont {Ross}},\ }\href@noop {} {\bibfield
  {journal} {\bibinfo  {journal} {J. Phys. Chem. Solids}\ }\textbf {\bibinfo
  {volume} {37}},\ \bibinfo {pages} {1135} (\bibinfo {year}
  {1976})}\BibitemShut {NoStop}%
\bibitem [{\citenamefont {Howard}\ \emph {et~al.}(1978)\citenamefont {Howard},
  \citenamefont {Waddington},\ and\ \citenamefont
  {Wright}}]{howard1978vibrational}%
  \BibitemOpen
  \bibfield  {author} {\bibinfo {author} {\bibfnamefont {J.}~\bibnamefont
  {Howard}}, \bibinfo {author} {\bibfnamefont {T.~C.}\ \bibnamefont
  {Waddington}}, \ and\ \bibinfo {author} {\bibfnamefont {C.~J.}\ \bibnamefont
  {Wright}},\ }\href@noop {} {\bibfield  {journal} {\bibinfo  {journal} {Chem.
  Phys. Lett.}\ }\textbf {\bibinfo {volume} {56}},\ \bibinfo {pages} {258}
  (\bibinfo {year} {1978})}\BibitemShut {NoStop}%
\bibitem [{\citenamefont {Rush}\ \emph {et~al.}(1984)\citenamefont {Rush},
  \citenamefont {Rowe},\ and\ \citenamefont {Richter}}]{rush1984direct}%
  \BibitemOpen
  \bibfield  {author} {\bibinfo {author} {\bibfnamefont {J.}~\bibnamefont
  {Rush}}, \bibinfo {author} {\bibfnamefont {J.}~\bibnamefont {Rowe}}, \ and\
  \bibinfo {author} {\bibfnamefont {D.}~\bibnamefont {Richter}},\ }\href@noop
  {} {\bibfield  {journal} {\bibinfo  {journal} {Z. Phys. B Condens. Matter}\
  }\textbf {\bibinfo {volume} {55}},\ \bibinfo {pages} {283} (\bibinfo {year}
  {1984})}\BibitemShut {NoStop}%
\bibitem [{\citenamefont {Nicol}\ \emph {et~al.}(1987)\citenamefont {Nicol},
  \citenamefont {Rush},\ and\ \citenamefont {Kelley}}]{nicol1987neutron}%
  \BibitemOpen
  \bibfield  {author} {\bibinfo {author} {\bibfnamefont {J.~M.}\ \bibnamefont
  {Nicol}}, \bibinfo {author} {\bibfnamefont {J.~J.}\ \bibnamefont {Rush}}, \
  and\ \bibinfo {author} {\bibfnamefont {R.~D.}\ \bibnamefont {Kelley}},\
  }\href@noop {} {\bibfield  {journal} {\bibinfo  {journal} {Phys. Rev. B}\
  }\textbf {\bibinfo {volume} {36}},\ \bibinfo {pages} {9315} (\bibinfo {year}
  {1987})}\BibitemShut {NoStop}%
\bibitem [{\citenamefont {Nicol}\ \emph {et~al.}(1988)\citenamefont {Nicol},
  \citenamefont {Rush},\ and\ \citenamefont {Kelley}}]{nicol1988inelastic}%
  \BibitemOpen
  \bibfield  {author} {\bibinfo {author} {\bibfnamefont {J.~M.}\ \bibnamefont
  {Nicol}}, \bibinfo {author} {\bibfnamefont {J.~J.}\ \bibnamefont {Rush}}, \
  and\ \bibinfo {author} {\bibfnamefont {R.~D.}\ \bibnamefont {Kelley}},\
  }\href@noop {} {\bibfield  {journal} {\bibinfo  {journal} {Surf. Sci.}\
  }\textbf {\bibinfo {volume} {197}},\ \bibinfo {pages} {67} (\bibinfo {year}
  {1988})}\BibitemShut {NoStop}%
\bibitem [{\citenamefont {Kolesnikov}\ \emph {et~al.}(1991)\citenamefont
  {Kolesnikov}, \citenamefont {Natkaniec}, \citenamefont {Antonov},
  \citenamefont {Belash}, \citenamefont {Fedotov}, \citenamefont {Krawczyk},
  \citenamefont {Mayer},\ and\ \citenamefont
  {Ponyatovsky}}]{kolesnikov1991neutron}%
  \BibitemOpen
  \bibfield  {author} {\bibinfo {author} {\bibfnamefont {A.}~\bibnamefont
  {Kolesnikov}}, \bibinfo {author} {\bibfnamefont {I.}~\bibnamefont
  {Natkaniec}}, \bibinfo {author} {\bibfnamefont {V.}~\bibnamefont {Antonov}},
  \bibinfo {author} {\bibfnamefont {I.}~\bibnamefont {Belash}}, \bibinfo
  {author} {\bibfnamefont {V.}~\bibnamefont {Fedotov}}, \bibinfo {author}
  {\bibfnamefont {J.}~\bibnamefont {Krawczyk}}, \bibinfo {author}
  {\bibfnamefont {J.}~\bibnamefont {Mayer}}, \ and\ \bibinfo {author}
  {\bibfnamefont {E.}~\bibnamefont {Ponyatovsky}},\ }\href@noop {} {\bibfield
  {journal} {\bibinfo  {journal} {Physica B Condens. Matter}\ }\textbf
  {\bibinfo {volume} {174}},\ \bibinfo {pages} {257} (\bibinfo {year}
  {1991})}\BibitemShut {NoStop}%
\bibitem [{\citenamefont {Nakai}\ \emph {et~al.}(1992)\citenamefont {Nakai},
  \citenamefont {Akiba}, \citenamefont {Asano},\ and\ \citenamefont
  {Ikeda}}]{nakai1992neutron}%
  \BibitemOpen
  \bibfield  {author} {\bibinfo {author} {\bibfnamefont {Y.}~\bibnamefont
  {Nakai}}, \bibinfo {author} {\bibfnamefont {E.}~\bibnamefont {Akiba}},
  \bibinfo {author} {\bibfnamefont {H.}~\bibnamefont {Asano}}, \ and\ \bibinfo
  {author} {\bibfnamefont {S.}~\bibnamefont {Ikeda}},\ }\href@noop {}
  {\bibfield  {journal} {\bibinfo  {journal} {J. Phys. Soc. Jpn.}\ }\textbf
  {\bibinfo {volume} {61}},\ \bibinfo {pages} {1834} (\bibinfo {year}
  {1992})}\BibitemShut {NoStop}%
\bibitem [{\citenamefont {Stuhr}\ \emph {et~al.}(1995)\citenamefont {Stuhr},
  \citenamefont {Wipf}, \citenamefont {Udovic}, \citenamefont {Weissmuller},\
  and\ \citenamefont {Gleiter}}]{stuhr1995vibrational}%
  \BibitemOpen
  \bibfield  {author} {\bibinfo {author} {\bibfnamefont {U.}~\bibnamefont
  {Stuhr}}, \bibinfo {author} {\bibfnamefont {H.}~\bibnamefont {Wipf}},
  \bibinfo {author} {\bibfnamefont {T.}~\bibnamefont {Udovic}}, \bibinfo
  {author} {\bibfnamefont {J.}~\bibnamefont {Weissmuller}}, \ and\ \bibinfo
  {author} {\bibfnamefont {H.}~\bibnamefont {Gleiter}},\ }\href@noop {}
  {\bibfield  {journal} {\bibinfo  {journal} {J. Phys. Condens. Matter}\
  }\textbf {\bibinfo {volume} {7}},\ \bibinfo {pages} {219} (\bibinfo {year}
  {1995})}\BibitemShut {NoStop}%
\bibitem [{\citenamefont {Ross}\ \emph {et~al.}(1998)\citenamefont {Ross},
  \citenamefont {Antonov}, \citenamefont {Bokhenkov}, \citenamefont
  {Kolesnikov}, \citenamefont {Ponyatovsky},\ and\ \citenamefont
  {Tomkinson}}]{ross1998strong}%
  \BibitemOpen
  \bibfield  {author} {\bibinfo {author} {\bibfnamefont {D.~K.}\ \bibnamefont
  {Ross}}, \bibinfo {author} {\bibfnamefont {V.~E.}\ \bibnamefont {Antonov}},
  \bibinfo {author} {\bibfnamefont {E.~L.}\ \bibnamefont {Bokhenkov}}, \bibinfo
  {author} {\bibfnamefont {A.~I.}\ \bibnamefont {Kolesnikov}}, \bibinfo
  {author} {\bibfnamefont {E.~G.}\ \bibnamefont {Ponyatovsky}}, \ and\ \bibinfo
  {author} {\bibfnamefont {J.}~\bibnamefont {Tomkinson}},\ }\href@noop {}
  {\bibfield  {journal} {\bibinfo  {journal} {Phys. Rev. B}\ }\textbf {\bibinfo
  {volume} {58}},\ \bibinfo {pages} {2591} (\bibinfo {year}
  {1998})}\BibitemShut {NoStop}%
\bibitem [{\citenamefont {Kemali}\ \emph {et~al.}(2000)\citenamefont {Kemali},
  \citenamefont {Totolici}, \citenamefont {Ross},\ and\ \citenamefont
  {Morrison}}]{kemali2000inelastic}%
  \BibitemOpen
  \bibfield  {author} {\bibinfo {author} {\bibfnamefont {M.}~\bibnamefont
  {Kemali}}, \bibinfo {author} {\bibfnamefont {J.~E.}\ \bibnamefont
  {Totolici}}, \bibinfo {author} {\bibfnamefont {D.~K.}\ \bibnamefont {Ross}},
  \ and\ \bibinfo {author} {\bibfnamefont {I.}~\bibnamefont {Morrison}},\
  }\href@noop {} {\bibfield  {journal} {\bibinfo  {journal} {Phys. Rev. Lett.}\
  }\textbf {\bibinfo {volume} {84}},\ \bibinfo {pages} {1531} (\bibinfo {year}
  {2000})}\BibitemShut {NoStop}%
\bibitem [{\citenamefont {Heuser}\ \emph {et~al.}(2008)\citenamefont {Heuser},
  \citenamefont {Udovic},\ and\ \citenamefont {Ju}}]{heuser2008vibrational}%
  \BibitemOpen
  \bibfield  {author} {\bibinfo {author} {\bibfnamefont {B.~J.}\ \bibnamefont
  {Heuser}}, \bibinfo {author} {\bibfnamefont {T.~J.}\ \bibnamefont {Udovic}},
  \ and\ \bibinfo {author} {\bibfnamefont {H.}~\bibnamefont {Ju}},\ }\href@noop
  {} {\bibfield  {journal} {\bibinfo  {journal} {Phys. Rev. B}\ }\textbf
  {\bibinfo {volume} {78}},\ \bibinfo {pages} {214101} (\bibinfo {year}
  {2008})}\BibitemShut {NoStop}%
\bibitem [{\citenamefont {Heuser}\ and\ \citenamefont
  {Ju}(2011)}]{heuser2011small}%
  \BibitemOpen
  \bibfield  {author} {\bibinfo {author} {\bibfnamefont {B.~J.}\ \bibnamefont
  {Heuser}}\ and\ \bibinfo {author} {\bibfnamefont {H.}~\bibnamefont {Ju}},\
  }\href@noop {} {\bibfield  {journal} {\bibinfo  {journal} {Phys. Rev. B}\
  }\textbf {\bibinfo {volume} {83}},\ \bibinfo {pages} {094103} (\bibinfo
  {year} {2011})}\BibitemShut {NoStop}%
\bibitem [{\citenamefont {Ju}\ \emph {et~al.}(2011)\citenamefont {Ju},
  \citenamefont {Heuser}, \citenamefont {Abernathy},\ and\ \citenamefont
  {Udovic}}]{ju2011comparison}%
  \BibitemOpen
  \bibfield  {author} {\bibinfo {author} {\bibfnamefont {H.}~\bibnamefont
  {Ju}}, \bibinfo {author} {\bibfnamefont {B.~J.}\ \bibnamefont {Heuser}},
  \bibinfo {author} {\bibfnamefont {D.~L.}\ \bibnamefont {Abernathy}}, \ and\
  \bibinfo {author} {\bibfnamefont {T.~J.}\ \bibnamefont {Udovic}},\
  }\href@noop {} {\bibfield  {journal} {\bibinfo  {journal} {Nucl. Instrum.
  Methods Phys. Res. A: Accel. Spectrom. Detect. Assoc. Equip.}\ }\textbf
  {\bibinfo {volume} {654}},\ \bibinfo {pages} {522} (\bibinfo {year}
  {2011})}\BibitemShut {NoStop}%
\bibitem [{\citenamefont {Heuser}\ \emph {et~al.}(2014)\citenamefont {Heuser},
  \citenamefont {Trinkle}, \citenamefont {Jalarvo}, \citenamefont {Serio},
  \citenamefont {Schiavone}, \citenamefont {Mamontov},\ and\ \citenamefont
  {Tyagi}}]{heuser2014direct}%
  \BibitemOpen
  \bibfield  {author} {\bibinfo {author} {\bibfnamefont {B.~J.}\ \bibnamefont
  {Heuser}}, \bibinfo {author} {\bibfnamefont {D.~R.}\ \bibnamefont {Trinkle}},
  \bibinfo {author} {\bibfnamefont {N.}~\bibnamefont {Jalarvo}}, \bibinfo
  {author} {\bibfnamefont {J.}~\bibnamefont {Serio}}, \bibinfo {author}
  {\bibfnamefont {E.~J.}\ \bibnamefont {Schiavone}}, \bibinfo {author}
  {\bibfnamefont {E.}~\bibnamefont {Mamontov}}, \ and\ \bibinfo {author}
  {\bibfnamefont {M.}~\bibnamefont {Tyagi}},\ }\href@noop {} {\bibfield
  {journal} {\bibinfo  {journal} {Phys. Rev. Lett.}\ }\textbf {\bibinfo
  {volume} {113}},\ \bibinfo {pages} {025504} (\bibinfo {year}
  {2014})}\BibitemShut {NoStop}%
\bibitem [{\citenamefont {Kofu}\ \emph {et~al.}(2016)\citenamefont {Kofu},
  \citenamefont {Hashimoto}, \citenamefont {Akiba}, \citenamefont {Kobayashi},
  \citenamefont {Kitagawa}, \citenamefont {Tyagi}, \citenamefont {Faraone},
  \citenamefont {Copley}, \citenamefont {Lohstroh},\ and\ \citenamefont
  {Yamamuro}}]{kofu2016hydrogen}%
  \BibitemOpen
  \bibfield  {author} {\bibinfo {author} {\bibfnamefont {M.}~\bibnamefont
  {Kofu}}, \bibinfo {author} {\bibfnamefont {N.}~\bibnamefont {Hashimoto}},
  \bibinfo {author} {\bibfnamefont {H.}~\bibnamefont {Akiba}}, \bibinfo
  {author} {\bibfnamefont {H.}~\bibnamefont {Kobayashi}}, \bibinfo {author}
  {\bibfnamefont {H.}~\bibnamefont {Kitagawa}}, \bibinfo {author}
  {\bibfnamefont {M.}~\bibnamefont {Tyagi}}, \bibinfo {author} {\bibfnamefont
  {A.}~\bibnamefont {Faraone}}, \bibinfo {author} {\bibfnamefont {J.~R.~D.}\
  \bibnamefont {Copley}}, \bibinfo {author} {\bibfnamefont {W.}~\bibnamefont
  {Lohstroh}}, \ and\ \bibinfo {author} {\bibfnamefont {O.}~\bibnamefont
  {Yamamuro}},\ }\href@noop {} {\bibfield  {journal} {\bibinfo  {journal}
  {Phys. Rev. B}\ }\textbf {\bibinfo {volume} {94}},\ \bibinfo {pages} {064303}
  (\bibinfo {year} {2016})}\BibitemShut {NoStop}%
\bibitem [{\citenamefont {Kofu}\ \emph {et~al.}(2017)\citenamefont {Kofu},
  \citenamefont {Hashimoto}, \citenamefont {Akiba}, \citenamefont {Kobayashi},
  \citenamefont {Kitagawa}, \citenamefont {Iida}, \citenamefont {Nakamura},\
  and\ \citenamefont {Yamamuro}}]{kofu2017vibrational}%
  \BibitemOpen
  \bibfield  {author} {\bibinfo {author} {\bibfnamefont {M.}~\bibnamefont
  {Kofu}}, \bibinfo {author} {\bibfnamefont {N.}~\bibnamefont {Hashimoto}},
  \bibinfo {author} {\bibfnamefont {H.}~\bibnamefont {Akiba}}, \bibinfo
  {author} {\bibfnamefont {H.}~\bibnamefont {Kobayashi}}, \bibinfo {author}
  {\bibfnamefont {H.}~\bibnamefont {Kitagawa}}, \bibinfo {author}
  {\bibfnamefont {K.}~\bibnamefont {Iida}}, \bibinfo {author} {\bibfnamefont
  {M.}~\bibnamefont {Nakamura}}, \ and\ \bibinfo {author} {\bibfnamefont
  {O.}~\bibnamefont {Yamamuro}},\ }\href@noop {} {\bibfield  {journal}
  {\bibinfo  {journal} {Phys. Rev. B}\ }\textbf {\bibinfo {volume} {96}},\
  \bibinfo {pages} {054304} (\bibinfo {year} {2017})}\BibitemShut {NoStop}%
\bibitem [{\citenamefont {Kofu}\ and\ \citenamefont
  {Yamamuro}(2020)}]{kofu2020dynamics}%
  \BibitemOpen
  \bibfield  {author} {\bibinfo {author} {\bibfnamefont {M.}~\bibnamefont
  {Kofu}}\ and\ \bibinfo {author} {\bibfnamefont {O.}~\bibnamefont
  {Yamamuro}},\ }\href@noop {} {\bibfield  {journal} {\bibinfo  {journal}
  {Journal of the Physical Society of Japan}\ }\textbf {\bibinfo {volume}
  {89}},\ \bibinfo {pages} {051002} (\bibinfo {year} {2020})}\BibitemShut
  {NoStop}%
\bibitem [{\citenamefont {Otomo}\ \emph {et~al.}(2020)\citenamefont {Otomo},
  \citenamefont {Ikeda},\ and\ \citenamefont {Honda}}]{otomo2020structural}%
  \BibitemOpen
  \bibfield  {author} {\bibinfo {author} {\bibfnamefont {T.}~\bibnamefont
  {Otomo}}, \bibinfo {author} {\bibfnamefont {K.}~\bibnamefont {Ikeda}}, \ and\
  \bibinfo {author} {\bibfnamefont {T.}~\bibnamefont {Honda}},\ }\href@noop {}
  {\bibfield  {journal} {\bibinfo  {journal} {J. Phys. Soc. Jpn.}\ }\textbf
  {\bibinfo {volume} {89}},\ \bibinfo {pages} {051001} (\bibinfo {year}
  {2020})}\BibitemShut {NoStop}%
\bibitem [{\citenamefont {Antonov}\ \emph {et~al.}(2022)\citenamefont
  {Antonov}, \citenamefont {Fedotov}, \citenamefont {Ivanov}, \citenamefont
  {Kolesnikov}, \citenamefont {Kuzovnikov}, \citenamefont {Tkacz},\ and\
  \citenamefont {Yartys}}]{antonov2022lattice}%
  \BibitemOpen
  \bibfield  {author} {\bibinfo {author} {\bibfnamefont {V.~E.}\ \bibnamefont
  {Antonov}}, \bibinfo {author} {\bibfnamefont {V.~K.}\ \bibnamefont
  {Fedotov}}, \bibinfo {author} {\bibfnamefont {A.~S.}\ \bibnamefont {Ivanov}},
  \bibinfo {author} {\bibfnamefont {A.~I.}\ \bibnamefont {Kolesnikov}},
  \bibinfo {author} {\bibfnamefont {M.~A.}\ \bibnamefont {Kuzovnikov}},
  \bibinfo {author} {\bibfnamefont {M.}~\bibnamefont {Tkacz}}, \ and\ \bibinfo
  {author} {\bibfnamefont {V.~A.}\ \bibnamefont {Yartys}},\ }\href@noop {}
  {\bibfield  {journal} {\bibinfo  {journal} {J. Alloys Compd.}\ }\textbf
  {\bibinfo {volume} {905}},\ \bibinfo {pages} {164208} (\bibinfo {year}
  {2022})}\BibitemShut {NoStop}%
\bibitem [{\citenamefont {Rahman}\ \emph {et~al.}(1976)\citenamefont {Rahman},
  \citenamefont {Sk{\"o}ld}, \citenamefont {Pelizzari}, \citenamefont {Sinha},\
  and\ \citenamefont {Flotow}}]{rahman1976phonon}%
  \BibitemOpen
  \bibfield  {author} {\bibinfo {author} {\bibfnamefont {A.}~\bibnamefont
  {Rahman}}, \bibinfo {author} {\bibfnamefont {K.}~\bibnamefont {Sk{\"o}ld}},
  \bibinfo {author} {\bibfnamefont {C.}~\bibnamefont {Pelizzari}}, \bibinfo
  {author} {\bibfnamefont {S.}~\bibnamefont {Sinha}}, \ and\ \bibinfo {author}
  {\bibfnamefont {H.}~\bibnamefont {Flotow}},\ }\href@noop {} {\bibfield
  {journal} {\bibinfo  {journal} {Phys. Rev. B}\ }\textbf {\bibinfo {volume}
  {14}},\ \bibinfo {pages} {3630} (\bibinfo {year} {1976})}\BibitemShut
  {NoStop}%
\bibitem [{\citenamefont {Gillan}(1986)}]{gillan1986simulation}%
  \BibitemOpen
  \bibfield  {author} {\bibinfo {author} {\bibfnamefont {M.}~\bibnamefont
  {Gillan}},\ }\href@noop {} {\bibfield  {journal} {\bibinfo  {journal} {J.
  Phys. C Solid State Phys.}\ }\textbf {\bibinfo {volume} {19}},\ \bibinfo
  {pages} {6169} (\bibinfo {year} {1986})}\BibitemShut {NoStop}%
\bibitem [{\citenamefont {Salomons}(1990)}]{salomons1990lattice}%
  \BibitemOpen
  \bibfield  {author} {\bibinfo {author} {\bibfnamefont {E.}~\bibnamefont
  {Salomons}},\ }\href@noop {} {\bibfield  {journal} {\bibinfo  {journal} {J.
  Phys. Condens. Matter}\ }\textbf {\bibinfo {volume} {2}},\ \bibinfo {pages}
  {845} (\bibinfo {year} {1990})}\BibitemShut {NoStop}%
\bibitem [{\citenamefont {Els{\"a}sser}\ \emph {et~al.}(1991)\citenamefont
  {Els{\"a}sser}, \citenamefont {Ho}, \citenamefont {Chan},\ and\ \citenamefont
  {F{\"a}hnle}}]{elsasser1991vibrational}%
  \BibitemOpen
  \bibfield  {author} {\bibinfo {author} {\bibfnamefont {C.}~\bibnamefont
  {Els{\"a}sser}}, \bibinfo {author} {\bibfnamefont {K.~M.}\ \bibnamefont
  {Ho}}, \bibinfo {author} {\bibfnamefont {C.~T.}\ \bibnamefont {Chan}}, \ and\
  \bibinfo {author} {\bibfnamefont {M.}~\bibnamefont {F{\"a}hnle}},\
  }\href@noop {} {\bibfield  {journal} {\bibinfo  {journal} {Phys. Rev. B}\
  }\textbf {\bibinfo {volume} {44}},\ \bibinfo {pages} {10377} (\bibinfo {year}
  {1991})}\BibitemShut {NoStop}%
\bibitem [{\citenamefont {Li}\ and\ \citenamefont
  {Wahnstr{\"o}m}(1992)}]{li1992molecular}%
  \BibitemOpen
  \bibfield  {author} {\bibinfo {author} {\bibfnamefont {Y.}~\bibnamefont
  {Li}}\ and\ \bibinfo {author} {\bibfnamefont {G.}~\bibnamefont
  {Wahnstr{\"o}m}},\ }\href@noop {} {\bibfield  {journal} {\bibinfo  {journal}
  {Phys. Rev. B}\ }\textbf {\bibinfo {volume} {46}},\ \bibinfo {pages} {14528}
  (\bibinfo {year} {1992})}\BibitemShut {NoStop}%
\bibitem [{\citenamefont {Trinkle}\ \emph {et~al.}(2011)\citenamefont
  {Trinkle}, \citenamefont {Ju}, \citenamefont {Heuser},\ and\ \citenamefont
  {Udovic}}]{trinkle2011nanoscale}%
  \BibitemOpen
  \bibfield  {author} {\bibinfo {author} {\bibfnamefont {D.~R.}\ \bibnamefont
  {Trinkle}}, \bibinfo {author} {\bibfnamefont {H.}~\bibnamefont {Ju}},
  \bibinfo {author} {\bibfnamefont {B.~J.}\ \bibnamefont {Heuser}}, \ and\
  \bibinfo {author} {\bibfnamefont {T.~J.}\ \bibnamefont {Udovic}},\
  }\href@noop {} {\bibfield  {journal} {\bibinfo  {journal} {Phys. Rev. B}\
  }\textbf {\bibinfo {volume} {83}},\ \bibinfo {pages} {174116} (\bibinfo
  {year} {2011})}\BibitemShut {NoStop}%
\bibitem [{\citenamefont {Errea}\ \emph {et~al.}(2013)\citenamefont {Errea},
  \citenamefont {Calandra},\ and\ \citenamefont {Mauri}}]{errea2013first}%
  \BibitemOpen
  \bibfield  {author} {\bibinfo {author} {\bibfnamefont {I.}~\bibnamefont
  {Errea}}, \bibinfo {author} {\bibfnamefont {M.}~\bibnamefont {Calandra}}, \
  and\ \bibinfo {author} {\bibfnamefont {F.}~\bibnamefont {Mauri}},\
  }\href@noop {} {\bibfield  {journal} {\bibinfo  {journal} {Phys. Rev. Lett.}\
  }\textbf {\bibinfo {volume} {111}},\ \bibinfo {pages} {177002} (\bibinfo
  {year} {2013})}\BibitemShut {NoStop}%
\bibitem [{\citenamefont {Paulatto}\ \emph {et~al.}(2015)\citenamefont
  {Paulatto}, \citenamefont {Errea}, \citenamefont {Calandra},\ and\
  \citenamefont {Mauri}}]{paulatto2015first}%
  \BibitemOpen
  \bibfield  {author} {\bibinfo {author} {\bibfnamefont {L.}~\bibnamefont
  {Paulatto}}, \bibinfo {author} {\bibfnamefont {I.}~\bibnamefont {Errea}},
  \bibinfo {author} {\bibfnamefont {M.}~\bibnamefont {Calandra}}, \ and\
  \bibinfo {author} {\bibfnamefont {F.}~\bibnamefont {Mauri}},\ }\href@noop {}
  {\bibfield  {journal} {\bibinfo  {journal} {Phys. Rev. B}\ }\textbf {\bibinfo
  {volume} {91}},\ \bibinfo {pages} {054304} (\bibinfo {year}
  {2015})}\BibitemShut {NoStop}%
\bibitem [{\citenamefont {Caputo}\ and\ \citenamefont
  {Alavi}(2003)}]{caputo2003h}%
  \BibitemOpen
  \bibfield  {author} {\bibinfo {author} {\bibfnamefont {R.}~\bibnamefont
  {Caputo}}\ and\ \bibinfo {author} {\bibfnamefont {A.}~\bibnamefont {Alavi}},\
  }\href@noop {} {\bibfield  {journal} {\bibinfo  {journal} {Mol. Phys.}\
  }\textbf {\bibinfo {volume} {101}},\ \bibinfo {pages} {1781} (\bibinfo {year}
  {2003})}\BibitemShut {NoStop}%
\bibitem [{\citenamefont {Ishimoto}\ and\ \citenamefont
  {Koyama}(2018)}]{ishimoto2018theoretical}%
  \BibitemOpen
  \bibfield  {author} {\bibinfo {author} {\bibfnamefont {T.}~\bibnamefont
  {Ishimoto}}\ and\ \bibinfo {author} {\bibfnamefont {M.}~\bibnamefont
  {Koyama}},\ }\href@noop {} {\bibfield  {journal} {\bibinfo  {journal} {J.
  Chem. Phys.}\ }\textbf {\bibinfo {volume} {148}} (\bibinfo {year}
  {2018})}\BibitemShut {NoStop}%
\bibitem [{\citenamefont {Els{\"a}sser}\ \emph {et~al.}(1992)\citenamefont
  {Els{\"a}sser}, \citenamefont {Ho}, \citenamefont {Chan},\ and\ \citenamefont
  {Fahnle}}]{elsasser1992first}%
  \BibitemOpen
  \bibfield  {author} {\bibinfo {author} {\bibfnamefont {C.}~\bibnamefont
  {Els{\"a}sser}}, \bibinfo {author} {\bibfnamefont {K.~M.}\ \bibnamefont
  {Ho}}, \bibinfo {author} {\bibfnamefont {C.~T.}\ \bibnamefont {Chan}}, \ and\
  \bibinfo {author} {\bibfnamefont {M.}~\bibnamefont {Fahnle}},\ }\href@noop {}
  {\bibfield  {journal} {\bibinfo  {journal} {J. Phys. Condens. Matter}\
  }\textbf {\bibinfo {volume} {4}},\ \bibinfo {pages} {5207} (\bibinfo {year}
  {1992})}\BibitemShut {NoStop}%
\bibitem [{\citenamefont {Ozawa}\ \emph {et~al.}(2023)\citenamefont {Ozawa},
  \citenamefont {Nakanishi}, \citenamefont {Kato}, \citenamefont {Shimizu},
  \citenamefont {Hitosugi},\ and\ \citenamefont
  {Fukutani}}]{ozawa2023observation}%
  \BibitemOpen
  \bibfield  {author} {\bibinfo {author} {\bibfnamefont {T.}~\bibnamefont
  {Ozawa}}, \bibinfo {author} {\bibfnamefont {H.}~\bibnamefont {Nakanishi}},
  \bibinfo {author} {\bibfnamefont {K.}~\bibnamefont {Kato}}, \bibinfo {author}
  {\bibfnamefont {R.}~\bibnamefont {Shimizu}}, \bibinfo {author} {\bibfnamefont
  {T.}~\bibnamefont {Hitosugi}}, \ and\ \bibinfo {author} {\bibfnamefont
  {K.}~\bibnamefont {Fukutani}},\ }\href@noop {} {\bibfield  {journal}
  {\bibinfo  {journal} {J. Phys. Chem. Solids}\ }\textbf {\bibinfo {volume}
  {185}},\ \bibinfo {pages} {111741} (\bibinfo {year} {2023})}\BibitemShut
  {NoStop}%
\bibitem [{\citenamefont {Parrinello}\ and\ \citenamefont
  {Rahman}(1984)}]{parrinello1984study}%
  \BibitemOpen
  \bibfield  {author} {\bibinfo {author} {\bibfnamefont {M.}~\bibnamefont
  {Parrinello}}\ and\ \bibinfo {author} {\bibfnamefont {A.}~\bibnamefont
  {Rahman}},\ }\href@noop {} {\bibfield  {journal} {\bibinfo  {journal} {J.
  Chem. Phys.}\ }\textbf {\bibinfo {volume} {80}},\ \bibinfo {pages} {860}
  (\bibinfo {year} {1984})}\BibitemShut {NoStop}%
\bibitem [{\citenamefont {Tuckerman}\ \emph {et~al.}(1993)\citenamefont
  {Tuckerman}, \citenamefont {Berne}, \citenamefont {Martyna},\ and\
  \citenamefont {Klein}}]{tuckerman1993efficient}%
  \BibitemOpen
  \bibfield  {author} {\bibinfo {author} {\bibfnamefont {M.~E.}\ \bibnamefont
  {Tuckerman}}, \bibinfo {author} {\bibfnamefont {B.~J.}\ \bibnamefont
  {Berne}}, \bibinfo {author} {\bibfnamefont {G.~J.}\ \bibnamefont {Martyna}},
  \ and\ \bibinfo {author} {\bibfnamefont {M.~L.}\ \bibnamefont {Klein}},\
  }\href@noop {} {\bibfield  {journal} {\bibinfo  {journal} {J. Chem. Phys.}\
  }\textbf {\bibinfo {volume} {99}},\ \bibinfo {pages} {2796} (\bibinfo {year}
  {1993})}\BibitemShut {NoStop}%
\bibitem [{\citenamefont {Feynman}(1972)}]{feynman1972statistical}%
  \BibitemOpen
  \bibfield  {author} {\bibinfo {author} {\bibfnamefont {R.~P.}\ \bibnamefont
  {Feynman}},\ }\href@noop {} {\emph {\bibinfo {title} {Statistical Mechanics,
  A Set of Lectures, California, Institute of Technology}}}\ (\bibinfo
  {publisher} {WA Benjamin, Inc. Advanced Book Program Reading,
  Massachusetts},\ \bibinfo {year} {1972})\BibitemShut {NoStop}%
\bibitem [{\citenamefont {Feynman}\ \emph {et~al.}(2010)\citenamefont
  {Feynman}, \citenamefont {Hibbs},\ and\ \citenamefont
  {Styer}}]{feynman2010quantum}%
  \BibitemOpen
  \bibfield  {author} {\bibinfo {author} {\bibfnamefont {R.~P.}\ \bibnamefont
  {Feynman}}, \bibinfo {author} {\bibfnamefont {A.~R.}\ \bibnamefont {Hibbs}},
  \ and\ \bibinfo {author} {\bibfnamefont {D.~F.}\ \bibnamefont {Styer}},\
  }\href@noop {} {\emph {\bibinfo {title} {Quantum mechanics and path
  integrals}}}\ (\bibinfo  {publisher} {Courier Corporation},\ \bibinfo {year}
  {2010})\BibitemShut {NoStop}%
\bibitem [{\citenamefont {Schulman}(2012)}]{schulman2012techniques}%
  \BibitemOpen
  \bibfield  {author} {\bibinfo {author} {\bibfnamefont {L.~S.}\ \bibnamefont
  {Schulman}},\ }\href@noop {} {\emph {\bibinfo {title} {Techniques and
  applications of path integration}}}\ (\bibinfo  {publisher} {Courier
  Corporation},\ \bibinfo {year} {2012})\BibitemShut {NoStop}%
\bibitem [{\citenamefont {Chandler}\ and\ \citenamefont
  {Wolynes}(1981)}]{chandler1981exploiting}%
  \BibitemOpen
  \bibfield  {author} {\bibinfo {author} {\bibfnamefont {D.}~\bibnamefont
  {Chandler}}\ and\ \bibinfo {author} {\bibfnamefont {P.~G.}\ \bibnamefont
  {Wolynes}},\ }\href@noop {} {\bibfield  {journal} {\bibinfo  {journal} {J.
  Chem. Phys.}\ }\textbf {\bibinfo {volume} {74}},\ \bibinfo {pages} {4078}
  (\bibinfo {year} {1981})}\BibitemShut {NoStop}%
\bibitem [{\citenamefont {Marx}\ and\ \citenamefont
  {Hutter}(2009)}]{marx2009ab}%
  \BibitemOpen
  \bibfield  {author} {\bibinfo {author} {\bibfnamefont {D.}~\bibnamefont
  {Marx}}\ and\ \bibinfo {author} {\bibfnamefont {J.}~\bibnamefont {Hutter}},\
  }\href@noop {} {\emph {\bibinfo {title} {Ab initio molecular dynamics: basic
  theory and advanced methods}}}\ (\bibinfo  {publisher} {Cambridge University
  Press},\ \bibinfo {year} {2009})\BibitemShut {NoStop}%
\bibitem [{\citenamefont {Tuckerman}(2010)}]{tuckerman2010statistical}%
  \BibitemOpen
  \bibfield  {author} {\bibinfo {author} {\bibfnamefont {M.}~\bibnamefont
  {Tuckerman}},\ }\href@noop {} {\emph {\bibinfo {title} {Statistical
  mechanics: theory and molecular simulation}}}\ (\bibinfo  {publisher} {Oxford
  University Press},\ \bibinfo {year} {2010})\BibitemShut {NoStop}%
\bibitem [{\citenamefont {Shiga}(2018)}]{shiga2018path}%
  \BibitemOpen
  \bibfield  {author} {\bibinfo {author} {\bibfnamefont {M.}~\bibnamefont
  {Shiga}},\ }\href@noop {} {\bibfield  {journal} {\bibinfo  {journal}
  {Reference Module in Chemistry, Molecular Sciences and Chemical Engineering}\
  } (\bibinfo {year} {2018})}\BibitemShut {NoStop}%
\bibitem [{\citenamefont {Markland}\ and\ \citenamefont
  {Ceriotti}(2018)}]{markland2018nuclear}%
  \BibitemOpen
  \bibfield  {author} {\bibinfo {author} {\bibfnamefont {T.~E.}\ \bibnamefont
  {Markland}}\ and\ \bibinfo {author} {\bibfnamefont {M.}~\bibnamefont
  {Ceriotti}},\ }\href@noop {} {\bibfield  {journal} {\bibinfo  {journal} {Nat.
  Rev. Chem.}\ }\textbf {\bibinfo {volume} {2}},\ \bibinfo {pages} {0109}
  (\bibinfo {year} {2018})}\BibitemShut {NoStop}%
\bibitem [{\citenamefont {Thomsen}\ and\ \citenamefont
  {Shiga}(2022)}]{thomsen2022structures}%
  \BibitemOpen
  \bibfield  {author} {\bibinfo {author} {\bibfnamefont {B.}~\bibnamefont
  {Thomsen}}\ and\ \bibinfo {author} {\bibfnamefont {M.}~\bibnamefont
  {Shiga}},\ }\href@noop {} {\bibfield  {journal} {\bibinfo  {journal} {Phys.
  Chem. Chem. Phys.}\ }\textbf {\bibinfo {volume} {24}},\ \bibinfo {pages}
  {10851} (\bibinfo {year} {2022})}\BibitemShut {NoStop}%
\bibitem [{\citenamefont {Cao}\ and\ \citenamefont
  {Voth}(1994)}]{cao1994formulation-2}%
  \BibitemOpen
  \bibfield  {author} {\bibinfo {author} {\bibfnamefont {J.}~\bibnamefont
  {Cao}}\ and\ \bibinfo {author} {\bibfnamefont {G.~A.}\ \bibnamefont {Voth}},\
  }\href@noop {} {\bibfield  {journal} {\bibinfo  {journal} {J. Chem. Phys.}\
  }\textbf {\bibinfo {volume} {100}},\ \bibinfo {pages} {5106} (\bibinfo {year}
  {1994})}\BibitemShut {NoStop}%
\bibitem [{\citenamefont {Craig}\ and\ \citenamefont
  {Manolopoulos}(2004)}]{craig2004quantum}%
  \BibitemOpen
  \bibfield  {author} {\bibinfo {author} {\bibfnamefont {I.~R.}\ \bibnamefont
  {Craig}}\ and\ \bibinfo {author} {\bibfnamefont {D.~E.}\ \bibnamefont
  {Manolopoulos}},\ }\href@noop {} {\bibfield  {journal} {\bibinfo  {journal}
  {J. Chem. Phys.}\ }\textbf {\bibinfo {volume} {121}},\ \bibinfo {pages}
  {3368} (\bibinfo {year} {2004})}\BibitemShut {NoStop}%
\bibitem [{\citenamefont {Krajewski}\ and\ \citenamefont
  {M{\"u}ser}(2004)}]{krajewski2004quantum}%
  \BibitemOpen
  \bibfield  {author} {\bibinfo {author} {\bibfnamefont {F.~R.}\ \bibnamefont
  {Krajewski}}\ and\ \bibinfo {author} {\bibfnamefont {M.~H.}\ \bibnamefont
  {M{\"u}ser}},\ }\href@noop {} {\bibfield  {journal} {\bibinfo  {journal}
  {Phys. Rev. Lett.}\ }\textbf {\bibinfo {volume} {92}},\ \bibinfo {pages}
  {030601} (\bibinfo {year} {2004})}\BibitemShut {NoStop}%
\bibitem [{\citenamefont {Rossi}\ \emph {et~al.}(2014)\citenamefont {Rossi},
  \citenamefont {Ceriotti},\ and\ \citenamefont
  {Manolopoulos}}]{rossi2014remove}%
  \BibitemOpen
  \bibfield  {author} {\bibinfo {author} {\bibfnamefont {M.}~\bibnamefont
  {Rossi}}, \bibinfo {author} {\bibfnamefont {M.}~\bibnamefont {Ceriotti}}, \
  and\ \bibinfo {author} {\bibfnamefont {D.~E.}\ \bibnamefont {Manolopoulos}},\
  }\href@noop {} {\bibfield  {journal} {\bibinfo  {journal} {J. Chem. Phys.}\
  }\textbf {\bibinfo {volume} {140}},\ \bibinfo {pages} {234116} (\bibinfo
  {year} {2014})}\BibitemShut {NoStop}%
\bibitem [{\citenamefont {Liu}(2014)}]{liu2014path}%
  \BibitemOpen
  \bibfield  {author} {\bibinfo {author} {\bibfnamefont {J.}~\bibnamefont
  {Liu}},\ }\href@noop {} {\bibfield  {journal} {\bibinfo  {journal} {J. Chem.
  Phys.}\ }\textbf {\bibinfo {volume} {140}},\ \bibinfo {pages} {224107}
  (\bibinfo {year} {2014})}\BibitemShut {NoStop}%
\bibitem [{\citenamefont {Hele}\ \emph {et~al.}(2015)\citenamefont {Hele},
  \citenamefont {Willatt}, \citenamefont {Muolo},\ and\ \citenamefont
  {Althorpe}}]{hele2015boltzmann}%
  \BibitemOpen
  \bibfield  {author} {\bibinfo {author} {\bibfnamefont {T.~J.}\ \bibnamefont
  {Hele}}, \bibinfo {author} {\bibfnamefont {M.~J.}\ \bibnamefont {Willatt}},
  \bibinfo {author} {\bibfnamefont {A.}~\bibnamefont {Muolo}}, \ and\ \bibinfo
  {author} {\bibfnamefont {S.~C.}\ \bibnamefont {Althorpe}},\ }\href@noop {}
  {\bibfield  {journal} {\bibinfo  {journal} {J. Chem. Phys.}\ }\textbf
  {\bibinfo {volume} {142}} (\bibinfo {year} {2015})}\BibitemShut {NoStop}%
\bibitem [{\citenamefont {Cendagorta}\ \emph {et~al.}(2018)\citenamefont
  {Cendagorta}, \citenamefont {Ba{\v{c}}i{\'c}},\ and\ \citenamefont
  {Tuckerman}}]{cendagorta2018open}%
  \BibitemOpen
  \bibfield  {author} {\bibinfo {author} {\bibfnamefont {J.~R.}\ \bibnamefont
  {Cendagorta}}, \bibinfo {author} {\bibfnamefont {Z.}~\bibnamefont
  {Ba{\v{c}}i{\'c}}}, \ and\ \bibinfo {author} {\bibfnamefont {M.~E.}\
  \bibnamefont {Tuckerman}},\ }\href@noop {} {\bibfield  {journal} {\bibinfo
  {journal} {J. Chem. Phys.}\ }\textbf {\bibinfo {volume} {148}} (\bibinfo
  {year} {2018})}\BibitemShut {NoStop}%
\bibitem [{\citenamefont {Trenins}\ \emph {et~al.}(2019)\citenamefont
  {Trenins}, \citenamefont {Willatt},\ and\ \citenamefont
  {Althorpe}}]{trenins2019path}%
  \BibitemOpen
  \bibfield  {author} {\bibinfo {author} {\bibfnamefont {G.}~\bibnamefont
  {Trenins}}, \bibinfo {author} {\bibfnamefont {M.~J.}\ \bibnamefont
  {Willatt}}, \ and\ \bibinfo {author} {\bibfnamefont {S.~C.}\ \bibnamefont
  {Althorpe}},\ }\href@noop {} {\bibfield  {journal} {\bibinfo  {journal} {J.
  Chem. Phys.}\ }\textbf {\bibinfo {volume} {151}},\ \bibinfo {pages} {054109}
  (\bibinfo {year} {2019})}\BibitemShut {NoStop}%
\bibitem [{\citenamefont {Kapil}\ \emph {et~al.}(2020)\citenamefont {Kapil},
  \citenamefont {Wilkins}, \citenamefont {Lan},\ and\ \citenamefont
  {Ceriotti}}]{kapil2020inexpensive}%
  \BibitemOpen
  \bibfield  {author} {\bibinfo {author} {\bibfnamefont {V.}~\bibnamefont
  {Kapil}}, \bibinfo {author} {\bibfnamefont {D.~M.}\ \bibnamefont {Wilkins}},
  \bibinfo {author} {\bibfnamefont {J.}~\bibnamefont {Lan}}, \ and\ \bibinfo
  {author} {\bibfnamefont {M.}~\bibnamefont {Ceriotti}},\ }\href@noop {}
  {\bibfield  {journal} {\bibinfo  {journal} {J. Chem. Phys.}\ }\textbf
  {\bibinfo {volume} {152}},\ \bibinfo {pages} {124104} (\bibinfo {year}
  {2020})}\BibitemShut {NoStop}%
\bibitem [{\citenamefont {Hasegawa}(2023)}]{hasegawa2023nuclear}%
  \BibitemOpen
  \bibfield  {author} {\bibinfo {author} {\bibfnamefont {T.}~\bibnamefont
  {Hasegawa}},\ }\href@noop {} {\bibfield  {journal} {\bibinfo  {journal} {J.
  Phys. Chem. Lett.}\ }\textbf {\bibinfo {volume} {14}},\ \bibinfo {pages}
  {8043} (\bibinfo {year} {2023})}\BibitemShut {NoStop}%
\bibitem [{\citenamefont {Shiga}(2022)}]{shiga2022path}%
  \BibitemOpen
  \bibfield  {author} {\bibinfo {author} {\bibfnamefont {M.}~\bibnamefont
  {Shiga}},\ }\href@noop {} {\bibfield  {journal} {\bibinfo  {journal} {J.
  Comput. Chem.}\ }\textbf {\bibinfo {volume} {43}},\ \bibinfo {pages} {1864}
  (\bibinfo {year} {2022})}\BibitemShut {NoStop}%
\bibitem [{\citenamefont {Kimizuka}\ \emph {et~al.}(2018)\citenamefont
  {Kimizuka}, \citenamefont {Ogata},\ and\ \citenamefont
  {Shiga}}]{kimizuka2018mechanism}%
  \BibitemOpen
  \bibfield  {author} {\bibinfo {author} {\bibfnamefont {H.}~\bibnamefont
  {Kimizuka}}, \bibinfo {author} {\bibfnamefont {S.}~\bibnamefont {Ogata}}, \
  and\ \bibinfo {author} {\bibfnamefont {M.}~\bibnamefont {Shiga}},\
  }\href@noop {} {\bibfield  {journal} {\bibinfo  {journal} {Phys. Rev. B}\
  }\textbf {\bibinfo {volume} {97}},\ \bibinfo {pages} {014102} (\bibinfo
  {year} {2018})}\BibitemShut {NoStop}%
\bibitem [{\citenamefont {Kimizuka}\ \emph {et~al.}(2019)\citenamefont
  {Kimizuka}, \citenamefont {Ogata},\ and\ \citenamefont
  {Shiga}}]{kimizuka2019unraveling}%
  \BibitemOpen
  \bibfield  {author} {\bibinfo {author} {\bibfnamefont {H.}~\bibnamefont
  {Kimizuka}}, \bibinfo {author} {\bibfnamefont {S.}~\bibnamefont {Ogata}}, \
  and\ \bibinfo {author} {\bibfnamefont {M.}~\bibnamefont {Shiga}},\
  }\href@noop {} {\bibfield  {journal} {\bibinfo  {journal} {Phys. Rev. B}\
  }\textbf {\bibinfo {volume} {100}},\ \bibinfo {pages} {024104} (\bibinfo
  {year} {2019})}\BibitemShut {NoStop}%
\bibitem [{\citenamefont {Kimizuka}\ and\ \citenamefont
  {Shiga}(2021)}]{kimizuka2021two}%
  \BibitemOpen
  \bibfield  {author} {\bibinfo {author} {\bibfnamefont {H.}~\bibnamefont
  {Kimizuka}}\ and\ \bibinfo {author} {\bibfnamefont {M.}~\bibnamefont
  {Shiga}},\ }\href@noop {} {\bibfield  {journal} {\bibinfo  {journal} {Phys.
  Rev. Mater.}\ }\textbf {\bibinfo {volume} {5}},\ \bibinfo {pages} {065406}
  (\bibinfo {year} {2021})}\BibitemShut {NoStop}%
\bibitem [{\citenamefont {Kwon}\ \emph {et~al.}(2023)\citenamefont {Kwon},
  \citenamefont {Shiga}, \citenamefont {Kimizuka},\ and\ \citenamefont
  {Oda}}]{kwon2023accurate}%
  \BibitemOpen
  \bibfield  {author} {\bibinfo {author} {\bibfnamefont {H.}~\bibnamefont
  {Kwon}}, \bibinfo {author} {\bibfnamefont {M.}~\bibnamefont {Shiga}},
  \bibinfo {author} {\bibfnamefont {H.}~\bibnamefont {Kimizuka}}, \ and\
  \bibinfo {author} {\bibfnamefont {T.}~\bibnamefont {Oda}},\ }\href@noop {}
  {\bibfield  {journal} {\bibinfo  {journal} {Acta Mater.}\ }\textbf {\bibinfo
  {volume} {247}},\ \bibinfo {pages} {118739} (\bibinfo {year}
  {2023})}\BibitemShut {NoStop}%
\bibitem [{\citenamefont {Witt}\ \emph {et~al.}(2009)\citenamefont {Witt},
  \citenamefont {Ivanov}, \citenamefont {Shiga}, \citenamefont {Forbert},\ and\
  \citenamefont {Marx}}]{witt2009applicability}%
  \BibitemOpen
  \bibfield  {author} {\bibinfo {author} {\bibfnamefont {A.}~\bibnamefont
  {Witt}}, \bibinfo {author} {\bibfnamefont {S.~D.}\ \bibnamefont {Ivanov}},
  \bibinfo {author} {\bibfnamefont {M.}~\bibnamefont {Shiga}}, \bibinfo
  {author} {\bibfnamefont {H.}~\bibnamefont {Forbert}}, \ and\ \bibinfo
  {author} {\bibfnamefont {D.}~\bibnamefont {Marx}},\ }\href@noop {} {\bibfield
   {journal} {\bibinfo  {journal} {J. Chem. Phys.}\ }\textbf {\bibinfo {volume}
  {130}},\ \bibinfo {pages} {194510} (\bibinfo {year} {2009})}\BibitemShut
  {NoStop}%
\bibitem [{\citenamefont {Ivanov}\ \emph {et~al.}(2010)\citenamefont {Ivanov},
  \citenamefont {Witt}, \citenamefont {Shiga},\ and\ \citenamefont
  {Marx}}]{ivanov2010artificial}%
  \BibitemOpen
  \bibfield  {author} {\bibinfo {author} {\bibfnamefont {S.~D.}\ \bibnamefont
  {Ivanov}}, \bibinfo {author} {\bibfnamefont {A.}~\bibnamefont {Witt}},
  \bibinfo {author} {\bibfnamefont {M.}~\bibnamefont {Shiga}}, \ and\ \bibinfo
  {author} {\bibfnamefont {D.}~\bibnamefont {Marx}},\ }\href@noop {} {\bibfield
   {journal} {\bibinfo  {journal} {J. Chem. Phys.}\ }\textbf {\bibinfo {volume}
  {132}},\ \bibinfo {pages} {031101} (\bibinfo {year} {2010})}\BibitemShut
  {NoStop}%
\bibitem [{\citenamefont {Miura}\ \emph {et~al.}(1999)\citenamefont {Miura},
  \citenamefont {Okazaki},\ and\ \citenamefont {Kinugawa}}]{miura1999path}%
  \BibitemOpen
  \bibfield  {author} {\bibinfo {author} {\bibfnamefont {S.}~\bibnamefont
  {Miura}}, \bibinfo {author} {\bibfnamefont {S.}~\bibnamefont {Okazaki}}, \
  and\ \bibinfo {author} {\bibfnamefont {K.}~\bibnamefont {Kinugawa}},\
  }\href@noop {} {\bibfield  {journal} {\bibinfo  {journal} {J. Chem. Phys.}\
  }\textbf {\bibinfo {volume} {110}},\ \bibinfo {pages} {4523} (\bibinfo {year}
  {1999})}\BibitemShut {NoStop}%
\bibitem [{\citenamefont {Craig}\ and\ \citenamefont
  {Manolopoulos}(2006)}]{craig2006inelastic}%
  \BibitemOpen
  \bibfield  {author} {\bibinfo {author} {\bibfnamefont {I.~R.}\ \bibnamefont
  {Craig}}\ and\ \bibinfo {author} {\bibfnamefont {D.~E.}\ \bibnamefont
  {Manolopoulos}},\ }\href@noop {} {\bibfield  {journal} {\bibinfo  {journal}
  {Chem. Phys.}\ }\textbf {\bibinfo {volume} {322}},\ \bibinfo {pages} {236}
  (\bibinfo {year} {2006})}\BibitemShut {NoStop}%
\bibitem [{\citenamefont {Lin}\ \emph {et~al.}(2003)\citenamefont {Lin},
  \citenamefont {Blanco},\ and\ \citenamefont {Goddard~III}}]{lin2003two}%
  \BibitemOpen
  \bibfield  {author} {\bibinfo {author} {\bibfnamefont {S.-T.}\ \bibnamefont
  {Lin}}, \bibinfo {author} {\bibfnamefont {M.}~\bibnamefont {Blanco}}, \ and\
  \bibinfo {author} {\bibfnamefont {W.~A.}\ \bibnamefont {Goddard~III}},\
  }\href@noop {} {\bibfield  {journal} {\bibinfo  {journal} {J. Chem. Phys.}\
  }\textbf {\bibinfo {volume} {119}},\ \bibinfo {pages} {11792} (\bibinfo
  {year} {2003})}\BibitemShut {NoStop}%
\bibitem [{\citenamefont {Behler}\ and\ \citenamefont
  {Parrinello}(2007)}]{behler2007generalized}%
  \BibitemOpen
  \bibfield  {author} {\bibinfo {author} {\bibfnamefont {J.}~\bibnamefont
  {Behler}}\ and\ \bibinfo {author} {\bibfnamefont {M.}~\bibnamefont
  {Parrinello}},\ }\href@noop {} {\bibfield  {journal} {\bibinfo  {journal}
  {Phys. Rev. Lett.}\ }\textbf {\bibinfo {volume} {98}},\ \bibinfo {pages}
  {146401} (\bibinfo {year} {2007})}\BibitemShut {NoStop}%
\bibitem [{\citenamefont {Behler}(2015)}]{behler2015constructing}%
  \BibitemOpen
  \bibfield  {author} {\bibinfo {author} {\bibfnamefont {J.}~\bibnamefont
  {Behler}},\ }\href@noop {} {\bibfield  {journal} {\bibinfo  {journal} {Int.
  J. Quant. Chem.}\ }\textbf {\bibinfo {volume} {115}},\ \bibinfo {pages}
  {1032} (\bibinfo {year} {2015})}\BibitemShut {NoStop}%
\bibitem [{\citenamefont {Behler}(2021)}]{behler2021four}%
  \BibitemOpen
  \bibfield  {author} {\bibinfo {author} {\bibfnamefont {J.}~\bibnamefont
  {Behler}},\ }\href@noop {} {\bibfield  {journal} {\bibinfo  {journal} {Chem.
  Rev.}\ }\textbf {\bibinfo {volume} {121}},\ \bibinfo {pages} {10037}
  (\bibinfo {year} {2021})}\BibitemShut {NoStop}%
\bibitem [{\citenamefont {Kimizuka}\ \emph {et~al.}(2022)\citenamefont
  {Kimizuka}, \citenamefont {Thomsen},\ and\ \citenamefont
  {Shiga}}]{kimizuka2022artificial}%
  \BibitemOpen
  \bibfield  {author} {\bibinfo {author} {\bibfnamefont {H.}~\bibnamefont
  {Kimizuka}}, \bibinfo {author} {\bibfnamefont {B.}~\bibnamefont {Thomsen}}, \
  and\ \bibinfo {author} {\bibfnamefont {M.}~\bibnamefont {Shiga}},\
  }\href@noop {} {\bibfield  {journal} {\bibinfo  {journal} {J. Phys. Energy}\
  }\textbf {\bibinfo {volume} {4}},\ \bibinfo {pages} {034004} (\bibinfo {year}
  {2022})}\BibitemShut {NoStop}%
\bibitem [{\citenamefont {Artrith}\ and\ \citenamefont
  {Urban}(2016)}]{artrith2016implementation}%
  \BibitemOpen
  \bibfield  {author} {\bibinfo {author} {\bibfnamefont {N.}~\bibnamefont
  {Artrith}}\ and\ \bibinfo {author} {\bibfnamefont {A.}~\bibnamefont
  {Urban}},\ }\href@noop {} {\bibfield  {journal} {\bibinfo  {journal} {Comput.
  Mater. Sci.}\ }\textbf {\bibinfo {volume} {114}},\ \bibinfo {pages} {135}
  (\bibinfo {year} {2016})}\BibitemShut {NoStop}%
\bibitem [{\citenamefont {Artrith}\ \emph {et~al.}(2017)\citenamefont
  {Artrith}, \citenamefont {Urban},\ and\ \citenamefont
  {Ceder}}]{artrith2017efficient}%
  \BibitemOpen
  \bibfield  {author} {\bibinfo {author} {\bibfnamefont {N.}~\bibnamefont
  {Artrith}}, \bibinfo {author} {\bibfnamefont {A.}~\bibnamefont {Urban}}, \
  and\ \bibinfo {author} {\bibfnamefont {G.}~\bibnamefont {Ceder}},\
  }\href@noop {} {\bibfield  {journal} {\bibinfo  {journal} {Phys. Rev. B}\
  }\textbf {\bibinfo {volume} {96}},\ \bibinfo {pages} {014112} (\bibinfo
  {year} {2017})}\BibitemShut {NoStop}%
\bibitem [{\citenamefont {Perdew}\ \emph {et~al.}(1996)\citenamefont {Perdew},
  \citenamefont {Burke},\ and\ \citenamefont
  {Ernzerhof}}]{perdew1996generalized}%
  \BibitemOpen
  \bibfield  {author} {\bibinfo {author} {\bibfnamefont {J.~P.}\ \bibnamefont
  {Perdew}}, \bibinfo {author} {\bibfnamefont {K.}~\bibnamefont {Burke}}, \
  and\ \bibinfo {author} {\bibfnamefont {M.}~\bibnamefont {Ernzerhof}},\
  }\href@noop {} {\bibfield  {journal} {\bibinfo  {journal} {Phys. Rev. Lett.}\
  }\textbf {\bibinfo {volume} {77}},\ \bibinfo {pages} {3865} (\bibinfo {year}
  {1996})}\BibitemShut {NoStop}%
\bibitem [{\citenamefont {Kresse}\ and\ \citenamefont
  {Furthm{\"u}ller}(1996)}]{kresse1996efficient}%
  \BibitemOpen
  \bibfield  {author} {\bibinfo {author} {\bibfnamefont {G.}~\bibnamefont
  {Kresse}}\ and\ \bibinfo {author} {\bibfnamefont {J.}~\bibnamefont
  {Furthm{\"u}ller}},\ }\href@noop {} {\bibfield  {journal} {\bibinfo
  {journal} {Phys. Rev. B}\ }\textbf {\bibinfo {volume} {54}},\ \bibinfo
  {pages} {11169} (\bibinfo {year} {1996})}\BibitemShut {NoStop}%
\bibitem [{\citenamefont {Nagai}\ \emph {et~al.}(2020)\citenamefont {Nagai},
  \citenamefont {Okumura}, \citenamefont {Kobayashi},\ and\ \citenamefont
  {Shiga}}]{nagai2020self}%
  \BibitemOpen
  \bibfield  {author} {\bibinfo {author} {\bibfnamefont {Y.}~\bibnamefont
  {Nagai}}, \bibinfo {author} {\bibfnamefont {M.}~\bibnamefont {Okumura}},
  \bibinfo {author} {\bibfnamefont {K.}~\bibnamefont {Kobayashi}}, \ and\
  \bibinfo {author} {\bibfnamefont {M.}~\bibnamefont {Shiga}},\ }\href@noop {}
  {\bibfield  {journal} {\bibinfo  {journal} {Phys. Rev. B}\ }\textbf {\bibinfo
  {volume} {102}},\ \bibinfo {pages} {041124(R)} (\bibinfo {year}
  {2020})}\BibitemShut {NoStop}%
\bibitem [{\citenamefont {Kobayashi}\ \emph {et~al.}(2021)\citenamefont
  {Kobayashi}, \citenamefont {Nagai}, \citenamefont {Itakura},\ and\
  \citenamefont {Shiga}}]{kobayashi2021self}%
  \BibitemOpen
  \bibfield  {author} {\bibinfo {author} {\bibfnamefont {K.}~\bibnamefont
  {Kobayashi}}, \bibinfo {author} {\bibfnamefont {Y.}~\bibnamefont {Nagai}},
  \bibinfo {author} {\bibfnamefont {M.}~\bibnamefont {Itakura}}, \ and\
  \bibinfo {author} {\bibfnamefont {M.}~\bibnamefont {Shiga}},\ }\href@noop {}
  {\bibfield  {journal} {\bibinfo  {journal} {J. Chem. Phys.}\ }\textbf
  {\bibinfo {volume} {155}} (\bibinfo {year} {2021})}\BibitemShut {NoStop}%
\bibitem [{\citenamefont {Shiga}\ \emph {et~al.}(2001)\citenamefont {Shiga},
  \citenamefont {Tachikawa},\ and\ \citenamefont {Miura}}]{shiga2001unified}%
  \BibitemOpen
  \bibfield  {author} {\bibinfo {author} {\bibfnamefont {M.}~\bibnamefont
  {Shiga}}, \bibinfo {author} {\bibfnamefont {M.}~\bibnamefont {Tachikawa}}, \
  and\ \bibinfo {author} {\bibfnamefont {S.}~\bibnamefont {Miura}},\
  }\href@noop {} {\bibfield  {journal} {\bibinfo  {journal} {J. Chem. Phys.}\
  }\textbf {\bibinfo {volume} {115}},\ \bibinfo {pages} {9149} (\bibinfo {year}
  {2001})}\BibitemShut {NoStop}%
\bibitem [{\citenamefont {Shiga}(2023)}]{shiga2023pimd}%
  \BibitemOpen
  \bibfield  {author} {\bibinfo {author} {\bibfnamefont {M.}~\bibnamefont
  {Shiga}},\ }\href@noop {} {\enquote {\bibinfo {title} {{PIMD}: An open-source
  software for parallel molecular simulations},}\ } (\bibinfo {year} {2023}),\
  \bibinfo {note}
  {https://ccse.jaea.go.jp/software/PIMD/index.en.html}\BibitemShut {NoStop}%
\bibitem [{\citenamefont {Ruiz-Barragan}\ \emph {et~al.}(2016)\citenamefont
  {Ruiz-Barragan}, \citenamefont {Ishimura},\ and\ \citenamefont
  {Shiga}}]{ruiz2016hierarchical}%
  \BibitemOpen
  \bibfield  {author} {\bibinfo {author} {\bibfnamefont {S.}~\bibnamefont
  {Ruiz-Barragan}}, \bibinfo {author} {\bibfnamefont {K.}~\bibnamefont
  {Ishimura}}, \ and\ \bibinfo {author} {\bibfnamefont {M.}~\bibnamefont
  {Shiga}},\ }\href@noop {} {\bibfield  {journal} {\bibinfo  {journal} {Chem.
  Phys. Lett.}\ }\textbf {\bibinfo {volume} {646}},\ \bibinfo {pages} {130}
  (\bibinfo {year} {2016})}\BibitemShut {NoStop}%
\bibitem [{\citenamefont {Borgschulte}\ \emph {et~al.}(2020)\citenamefont
  {Borgschulte}, \citenamefont {Terreni}, \citenamefont {Billeter},
  \citenamefont {Daemen}, \citenamefont {Cheng}, \citenamefont {Pandey},
  \citenamefont {{\L}odziana}, \citenamefont {Hemley},\ and\ \citenamefont
  {Ramirez-Cuesta}}]{borgschulte2020inelastic}%
  \BibitemOpen
  \bibfield  {author} {\bibinfo {author} {\bibfnamefont {A.}~\bibnamefont
  {Borgschulte}}, \bibinfo {author} {\bibfnamefont {J.}~\bibnamefont
  {Terreni}}, \bibinfo {author} {\bibfnamefont {E.}~\bibnamefont {Billeter}},
  \bibinfo {author} {\bibfnamefont {L.}~\bibnamefont {Daemen}}, \bibinfo
  {author} {\bibfnamefont {Y.}~\bibnamefont {Cheng}}, \bibinfo {author}
  {\bibfnamefont {A.}~\bibnamefont {Pandey}}, \bibinfo {author} {\bibfnamefont
  {Z.}~\bibnamefont {{\L}odziana}}, \bibinfo {author} {\bibfnamefont {R.~J.}\
  \bibnamefont {Hemley}}, \ and\ \bibinfo {author} {\bibfnamefont {A.~J.}\
  \bibnamefont {Ramirez-Cuesta}},\ }\href@noop {} {\bibfield  {journal}
  {\bibinfo  {journal} {Proc. Nat. Acad. Sci.}\ }\textbf {\bibinfo {volume}
  {117}},\ \bibinfo {pages} {4021} (\bibinfo {year} {2020})}\BibitemShut
  {NoStop}%
\bibitem [{\citenamefont {Akiba}\ \emph {et~al.}(2016)\citenamefont {Akiba},
  \citenamefont {Kofu}, \citenamefont {Kobayashi}, \citenamefont {Kitagawa},
  \citenamefont {Ikeda}, \citenamefont {Otomo},\ and\ \citenamefont
  {Yamamuro}}]{akiba2016nanometer}%
  \BibitemOpen
  \bibfield  {author} {\bibinfo {author} {\bibfnamefont {H.}~\bibnamefont
  {Akiba}}, \bibinfo {author} {\bibfnamefont {M.}~\bibnamefont {Kofu}},
  \bibinfo {author} {\bibfnamefont {H.}~\bibnamefont {Kobayashi}}, \bibinfo
  {author} {\bibfnamefont {H.}~\bibnamefont {Kitagawa}}, \bibinfo {author}
  {\bibfnamefont {K.}~\bibnamefont {Ikeda}}, \bibinfo {author} {\bibfnamefont
  {T.}~\bibnamefont {Otomo}}, \ and\ \bibinfo {author} {\bibfnamefont
  {O.}~\bibnamefont {Yamamuro}},\ }\href@noop {} {\bibfield  {journal}
  {\bibinfo  {journal} {J. Am. Chem. Soc.}\ }\textbf {\bibinfo {volume}
  {138}},\ \bibinfo {pages} {10238} (\bibinfo {year} {2016})}\BibitemShut
  {NoStop}%
\bibitem [{\citenamefont {Colbert}\ and\ \citenamefont
  {Miller}(1992)}]{colbert1992novel}%
  \BibitemOpen
  \bibfield  {author} {\bibinfo {author} {\bibfnamefont {D.~T.}\ \bibnamefont
  {Colbert}}\ and\ \bibinfo {author} {\bibfnamefont {W.~H.}\ \bibnamefont
  {Miller}},\ }\href@noop {} {\bibfield  {journal} {\bibinfo  {journal} {J.
  Chem. Phys.}\ }\textbf {\bibinfo {volume} {96}},\ \bibinfo {pages} {1982}
  (\bibinfo {year} {1992})}\BibitemShut {NoStop}%
\bibitem [{\citenamefont {P{\'e}rez}\ \emph {et~al.}(2009)\citenamefont
  {P{\'e}rez}, \citenamefont {Tuckerman},\ and\ \citenamefont
  {M{\"u}ser}}]{perez2009comparative}%
  \BibitemOpen
  \bibfield  {author} {\bibinfo {author} {\bibfnamefont {A.}~\bibnamefont
  {P{\'e}rez}}, \bibinfo {author} {\bibfnamefont {M.~E.}\ \bibnamefont
  {Tuckerman}}, \ and\ \bibinfo {author} {\bibfnamefont {M.~H.}\ \bibnamefont
  {M{\"u}ser}},\ }\href@noop {} {\bibfield  {journal} {\bibinfo  {journal} {J.
  Chem. Phys.}\ }\textbf {\bibinfo {volume} {130}},\ \bibinfo {pages} {184105}
  (\bibinfo {year} {2009})}\BibitemShut {NoStop}%
\end{thebibliography}%

\begin{table}[htbp]
\caption{Peak positions (with no lattice strain)}
\begin{center}
\begin{tabular}{ccccc}
\hline\hline
\hspace{1mm} Method \hspace{1mm} & \hspace{1mm} Site \hspace{1mm}
& \hspace{1mm}$T$ [K] \hspace{1mm} & \hspace{1mm} $k$ [\AA$^{-1}$] \hspace{1mm} & \hspace{1mm} Peaks [eV] \hspace{1mm} \\ 
\hline
Exptl.\cite{rush1984direct} &  & 295 &     & 69.0$\pm$0.5, 137$\pm$2 \\
BCMD & O & 300 & 0--4 & 67, - \\ 
BCMD & O & 300 & 4--8 & 68$^\ast$, 133$^\ast$ \\ 
BCMD & O & 300 & 8--12 & 68, 139 \\ 
BCMD & O & 200 & 4--8 & 68, 131 \\  
BCMD & O & 100 & 4--8 & 68, 133 \\  
BCMD & O &  75 & 0--4 & 69, 131 \\  
BCMD & O &  75 & 4--8 & 68, 133 \\  
BCMD & O &  75 & 8--12 & 68, 133 \\  
BCMD & O &  50 & 4--8 & 70, 140 \\ 
 CMD & O & 300 & 4--8 & 65, 125 \\ 
 CMD & O &  75 & 4--8 & 66, 132 \\ 
RPMD & O & 300 & 4--8 & 68$^\ast$, 135$^\ast$ \\ 
RPMD & O &  75 & 4--8 & 67, 153 \\ 
  MD & O & 300 & 4--8 & 53, - \\  
  MD & O & 200 & 4--8 & 53, - \\  
  MD & O & 100 & 4--8 & 47, - \\  
  MD & O &  75 & 4--8 & 45, 92 \\  
  MD & O &  50 & 4--8 & 42, 86 \\  
 HAR & O &   0 &      & 34.8 \\  %
 DVR & O &   0 &      & 70.3, 133.8 \\  %
BCMD & T &  75 & 0--4 & 127$^\ast$, 136$^\ast$ \\ 
BCMD & T &  75 & 4--8 & 127$^\ast$, 136$^\ast$ \\ 
BCMD & T &  75 & 8--12 & 127$^\ast$, 136$^\ast$ \\ 
 HAR & T &   0 &      & 125.8 \\ %
\hline
\end{tabular}
\end{center}
\label{tab1}
$^\ast$By fitting the peaks to two Lorentzian functions.
\end{table}
%
%
\newpage
%
%
\newpage
%
%
\newpage
%
%
\newpage
%
%
\newpage
%
%
\newpage
%
%
\newpage
%
%
\newpage
%
%
\newpage
%
%
\newpage
%
%
\newpage
%
%
\newpage
%
%
\newpage

\end{document}